\documentclass[longauth]{aa} 
\usepackage{graphicx}
\usepackage{natbib}
\usepackage{txfonts}
\begin{document}

   \title{IGAPS: the merged IPHAS and UVEX optical surveys of the Northern Galactic Plane}
   \titlerunning{IGAPS}

   \author{M. Mongui\'o
          \inst{1,2}
          \and
          R. Greimel\inst{3}\and
          J. E. Drew\inst{1,4}\and
          G. Barentsen\inst{1,5}\and
          P. J. Groot\inst{6,7,8,9}\and
          M. J. Irwin\inst{10} \and
          J. Casares\inst{11,12} \and
          B. T. G\"ansicke\inst{13}\and
P. J. Carter\inst{13,14}\and
J. M. Corral-Santana\inst{11,15} \and
N. P. Gentile-Fusillo\inst{13,15}\and
S. Greiss\inst{13} \and
L. M. van Haaften\inst{6,16}\and
M. Hollands\inst{13}\and
D. Jones\inst{11,12}\and
T. Kupfer\inst{6,17}\and
C. J. Manser\inst{13}\and
D. N. A. Murphy\inst{10}\and
A. F. McLeod\inst{6,16,18}\and
T. Oosting\inst{6}\and
Q. A. Parker\inst{19}\and
S. Pyrzas\inst{13,20}\and
P. Rodr\'iguez-Gil\inst{11,12}\and
J. van Roestel\inst{6,21}\and
S. Scaringi\inst{16}\and
P. Schellart\inst{6}\and
O. Toloza\inst{13}\and
O. Vaduvescu\inst{11,22}\and
L. van Spaandonk\inst{13,23}\and
K. Verbeek\inst{6}\and
N. J. Wright\inst{24}\and
J. Eisl\"offel\inst{25}\and
J. Fabregat\inst{26}\and 
A. Harris\inst{1}\and
R. A. H. Morris\inst{27}\and
S. Phillipps\inst{27}\and
R. Raddi\inst{13,28}\and
L. Sabin\inst{29}\and
Y. Unruh\inst{30}\and
J. S Vink\inst{31}\and
R. Wesson\inst{4} \and
A. Cardwell\inst{22,32}\and 
R. K. Cochrane\inst{22}\and
S. Doostmohammadi \inst{22,33}\and
T. Mocnik\inst{22}\and 
H. Stoev\inst{22}\and
L. Su\'arez-Andr\'es\inst{22}\and
V. Tudor \inst{22}\and
T. G. Wilson\inst{22}\and
T. J. Zegmott\inst{22}
          }

   \institute{
School of Physics, Astronomy \& Mathematics, University of Hertfordshire, Hatfield AL10 9AB, UK \\
\email{m.monguio@icc.ub.edu}\and
Institut d'Estudis Espacials de Catalunya, Universitat de Barcelona (ICC-UB), Mart\'i i Franqu\`es 1,
E-08028 Barcelona, Spain\and
IGAM, Institute of Physics, University of Graz, Universit{\"a}tsplatz 5/II, 8010 Graz, Austria\and
Department of Physics \& Astronomy, University College London, Gower Street, London WC1E 6BT, UK\and
Bay Area Environmental Research Institute, P.O. Box 25, Moffett Field, CA 94035, USA\and
Department of Astrophysics/IMAPP, Radboud University, P.O. Box 9010,
6500 GL Nijmegen, The Netherlands\and
Department of Astronomy, University of Cape Town, Private Bag X3, Rondebosch, 7701, South Africa\and
South African Astronomical Observatory, P.O. Box 9, Observatory,
7935, South Africa\and
The Inter-University Institute for Data Intensive Astronomy,
University of Cape Town, Private Bag X3, Rondebosch, 7701, South Africa\and
Institute of Astronomy, University of Cambridge, Madingley Road, Cambridge, CB3 0HA, UK\and
Instituto de Astrof\'isica de Canarias, E-38205 La Laguna, Tenerife, Spain\and
Departamento de Astrof\'isica, Universidad de La Laguna, E-38206 La Laguna, Tenerife, Spain\and
University of Warwick, Department of Physics, Gibbet Hill Road, Coventry, CV4 7AL, UK\and
Department of Earth and Planetary Sciences, University of California, Davis, One Shields Avenue, Davis, CA 95616, USA\and
European Southern Observatory (ESO), Av. Alonso de C\'ordova 3107, 7630355 Vitacura, Santiago, Chile\and
Department of Physics and Astronomy, Texas Tech University, PO Box 41051, Lubbock, TX 79409, USA\and
Kavli Institute for Theoretical Physics, University of California, Santa Barbara, CA 93106, USA\and
Department of Astronomy, University of California Berkeley, Berkeley, CA 94720, USA\and
The University of Hong Kong, Department of Physics, Hong Kong SAR, China\and
Hamad Bin Khalifa University (HBKU), Qatar Foundation, P.O. Box 5825, Doha, Qatar.\and
Division of Physics, Mathematics and Astronomy, California Institute of Technology, Pasadena, CA 91125, USA \and
Isaac Newton Group, Apartado de correos 321, E-38700 Santa Cruz de La Palma, Canary Islands, Spain\and
Mollerlyceum, 4611DX, Bergen op Zoom, The Netherlands\and
Astrophysics Group, Keele University, Keele, ST5 5BG, UK\and
Th\"uringer Landessternwarte, Sternwarte 5, D-07778 Tautenburg, Germany\and
Observatorio Astron\'omico, Universidad de Valencia, Calle Catedr\'atico Jos\'e Beltr\'an 2, 46980 Paterna, Spain \and
Astrophysics Group, School of Physics, University of Bristol, Tyndall Av, Bristol, BS8 1TL, UK\and
Dr. Remeis-Sternwarte, Friedrich Alexander Universit\"at Erlangen-N\"urnberg, 
Sternwartstr 7, D-96049 Bamberg, Germany\and
Universidad Nacional Aut\'onoma de M\'exico (UNAM), Instituto de Astronom\'ia, Km 103 Carretera Tijuana, Ensenada, Mexico\and
Department of Physics, Imperial College London, SW7 2AZ \and
Armagh Observatory and Planetarium, BT61 9DG, Armagh, UK\and
LBT Observatory, University of Arizona, 933 N. Cherry Ave, Tucson, AZ 85721-0009, U.S.A. \and
Department of Physics, Shahid Bahonar University of Kerman, Iran     }

   \date{Received December 17, 2019; accepted February 12, 2020}

  \abstract{The INT Galactic Plane Survey (IGAPS) is the merger of the optical photometric surveys, IPHAS and UVEX, based on data from the Isaac Newton Telescope (INT) obtained between 2003 and 2018. Here, we present the IGAPS point source catalogue. It contains 295.4 million rows providing photometry in the filters, $i$, $r$, narrow-band $H\alpha$, $g$ and $U_{RGO}$. The IGAPS footprint fills the Galactic coordinate range, $|b| < 5^{\circ}$ and $30^{\circ} < \ell < 215^{\circ}$.  A uniform calibration, referred to the Pan-STARRS system, is applied to $g$, $r$ and $i$, while the $H\alpha$ calibration is linked to $r$ and then is reconciled via field overlaps. The astrometry in all 5 bands has been recalculated on the Gaia DR2 frame.  Down to $i \sim 20$ mag. (Vega system), most stars are also detected in $g$, $r$ and $H\alpha$.  As exposures in the $r$ band were obtained within the IPHAS and UVEX surveys a few years apart, typically, the catalogue includes two distinct $r$ measures, $r_I$ and $r_U$.  The $r$ 10$\sigma$ limiting magnitude is $\sim$21, with median seeing 1.1 arcsec. Between $\sim$13th and $\sim$19th magnitudes in all bands, the photometry is internally reproducible to within 0.02 magnitudes.  Stars brighter than $r=19.5$ have been tested for narrow-band $H\alpha$ excess signalling line emission, and for variation exceeding $|r_I-r_U| = 0.2$ mag.  We find and flag 8292 candidate emission line stars and over 53000 variables (both at $>5\sigma$ confidence).  The 174-column catalogue will be available via CDS Strasbourg.}


   \maketitle
%

\section{Introduction}
\label{sec:intro}

The stellar and nebular content of the Galactic Plane continues to be a vitally important object of study as it offers the best available angular resolution to understand how galactic disc environments are built, interact and evolve over time.  The optical part of the electromagnetic spectrum remains an important window, particularly for characterising the properties of the disc's stellar content, as this is the range in which the Planck function maximum falls for most stars.  For studies of the interstellar medium, it is relevant that the optical is also the domain in which H$\alpha$, the strongest observable hydrogen emission line, is located.  This line is the outstanding tracer of ionized interstellar and circumstellar gas.  

In this era of digital surveys, there is a  growing menu of ground-based wide-field optical broad band surveys covering much of the sky, north and south \citep[SDSS, Pan-STARRS, APASS, DECaPS, Skymapper, see:][]{SDSS,Chambers16,APASS,DECaPS,Skymapper}.
Here we add to the menu by focusing on the dense star fields of the northern Milky Way, and by bring together for the first time, two Galactic Plane surveys that have each deployed a filter particularly well suited to searching for early and late phases of stellar evolution.   IPHAS \citep[The INT Photometric $H\alpha$ survey of the northern Galactic Plane, ][]{IPHAS} has incorporated imaging narrow-band $H\alpha$, while UVEX \citep[The UV-Excess survey of the northern Galactic Plane, ][]{UVEX} has included imaging using the Sloan-$u$-like $U_{RGO}$ filter.  In concept, these two surveys are the older siblings to VPHAS+, the survey covering the southern Galactic Plane and Bulge \citep{VPHAS}.  

A crucial and defining feature of the IPHAS and UVEX surveys is that their observing plans centered on contemporaneous observations in the full set of filters so as to achieve faithful colour information, immune to stellar variability on timescales longer than $\sim$10 minutes.  This characteristic is shared with the continuing Gaia mission \citep{2018A&A...616A...1G}.  Both IPHAS and UVEX were executed using the Wide Field Camera (WFC) on the Isaac Newton Telescope (INT) in La Palma. Together they form the largest scientific investigation so far undertaken at the INT, requiring more than 400 nights.

IPHAS and UVEX are respectively red-optical and blue-optical surveys. So that they could be linked together, photometrically, both surveys included the Sloan $r$ band in their filter sets.  This was also seen as an opportunity to look for evidence of both variability and measurable proper motion relative to a typical epoch difference of a few years.  We note that recent work by \cite{Scaringi2018} has already identified higher proper motion objects by comparing IPHAS $r$ and Gaia DR2 positions.  Here we will briefly consider the incidence of variability as revealed by the two epochs of IPHAS and UVEX $r$ band data.

This paper presents a calibration of the point source photometry in $r$/$i$/$H\alpha$ and $r$/$g$/$U_{RGO}$ collected by the IPHAS and UVEX surveys respectively, and their merger into a single catalogue recording data on almost 300 million objects. The broad band calibration is aligned with the Pan-STARRS photometric scale set by \cite{Magnier2013}, while the $H\alpha$ narrow band needs its own bespoke solution.  The final catalogue also benefits from a recalculation of the astrometry to place it into the Gaia DR2 astrometric reference frame. We note that in the case of IPHAS there have been two previous data releases \citep{IPHASIDR,IPHASDR2}. 
The last observations were UVEX exposures gathered in late 2018, bringing to an end a campaign on the INT that began with the first IPHAS observations in 2003.  The new acronym we adopt to represent the merged database is IGAPS, standing for "The INT Galactic Plane Survey".

Here we summarise the observing strategy, data pipelining and quality control shared between the two surveys in sections 2 and 3.  The way in which the astrometry is refitted in order to convert it from a 2MASS frame to that of Gaia DR2 is described in section 4. After this we turn to the global calibration of the UVEX $g$ and $r$ data alongside the $r$, $i$, and narrow-band $H\alpha$ data of the IPHAS survey (section 5).  All surveys have unwanted artefacts to deal with and we identify them and their mitigation in section 6.  Sections 7 and 8 describe the compilation of the photometric catalogue and its contents. Section 8 includes a comparison between IGAPS source counts and those of Gaia DR2 and Pan-STARRS. There is also a brief discussion of the 4 photometric colour-colour diagrams the catalogue supports.  In section 9, we report on a new selection of candidate emission line stars \citep[based on the $r-H\alpha$ versus $r-i$ diagram: see][]{Witham08}, and on the identification of stellar variables via the two epochs of $r$ observation contained within the catalogue.  Section 10 contains closing remarks.

\section{Observations and sky coverage}
\label{sec:Obs}

\begin{figure}\centering
 \resizebox{.98\hsize}{!}{\includegraphics{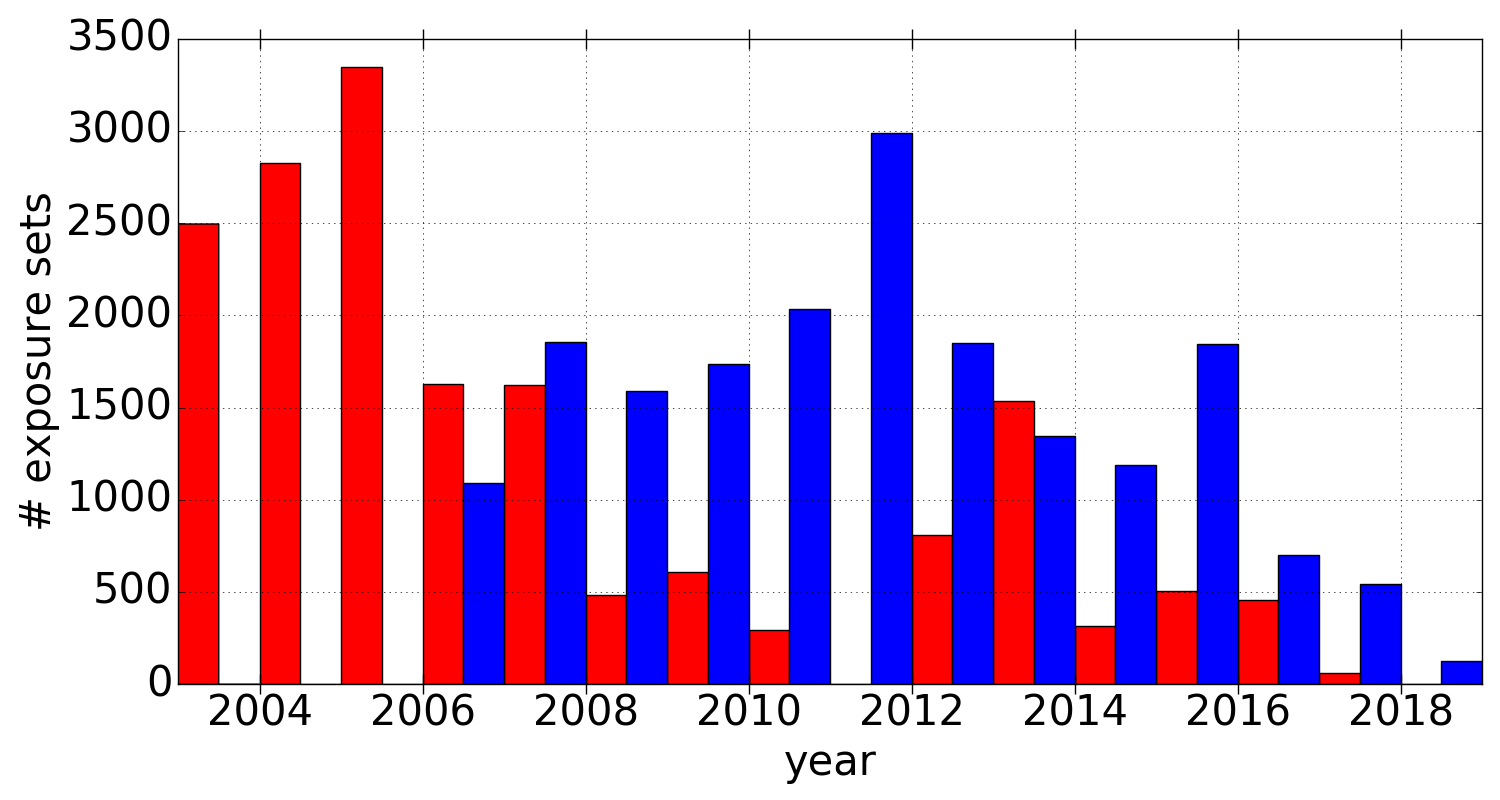}}
\caption{Number of 3-filter exposure sets obtained per year for IPHAS ($H\alpha$, $r$ and $i$, shown in red) and UVEX ($r$, $g$ and $U_{RGO}$, shown in blue).}
\label{fig:years}
\end{figure}

The survey observations were all obtained using the Wide Field Camera (WFC) mounted on the INT.  
IPHAS observations began in 2003, while the blue UVEX data gathering began in 2006. Most of the footprint had been covered once by the end of 2012, while the later observations mainly focused on repeats correcting for poor weather and other problems identified in quality control (see figure~\ref{fig:years}). 

Table~\ref{tab:basics} provides an overview of key features of the merged IGAPS survey.

\begin{table*}
\begin{tabular}{lll}
\hline
Property   &  Value   &  Comment \\
\hline
Telescope & 2.5-m Isaac Newton Telescope (INT) & \\
Instrument &  Wide Field Camera (WFC) & \\
Detectors  & Four 2048$\times$4100 pixel CCDs & \\
Pixel Scale & 0.33 arcsec pixel$^{-1}$ & \\
Filters & $i$, $H\alpha$, $r$, $g$, $U_{RGO}$ & 2 $r$ epochs available \\
Magnitude System & Vega & $m_{AB}$ provided as alternative \\
Exposure times (seconds) & $i$:10, $H_\alpha$:120, $r$:30, $g$:30, $U_{RGO}$:120 &  \\
Saturation magnitude & 12($i$), 12.5($H\alpha$), 13($r$), 14($g$) 14.5($U_{RGO}$) & \\
Limiting magnitude & 20.4($i$), 20.5($H\alpha$), 21.5($r$), 22.4($g$), 21.5($U_{RGO}$) & median 5$\sigma$ detection over the noise. \\
median PSF FWHM (arcsec) & 1.0($i$), 1.2($H\alpha$), 1.1($r$), 1.3($g$), 1.5($U_{RGO}$) & \\
Survey area & $\sim1860$ square degrees & \\ 
Footprint boundaries & $-5^{\circ} < b < +5^{\circ}$, $30^{\circ} < \ell < 215^{\circ}$& \\
Beginning/end dates of observations & August 2003 -- November 2018 & see Figure~\ref{fig:years} \\
\hline
\end{tabular}
\caption{Key properties of the merged IGAPS survey.}\label{tab:basics}
\end{table*}

The WFC is a 4-CCD mosaic arranged in an L shape with a pixel size of 0.33\,arcsec/pixel, and a total field of view spanning approximately 0.22 square degrees.
The five filters used\footnote{see http://catserver.ing.iac.es/filter/list.php?instrument=WFC} -- $U_{\rm RGO}$, $g$, $r$, $i$, $H\alpha$ -- have central wavelengths of 364.0, 484.6, 624.0, 774.3, 816.0\,nm respectively. Note that the $U_{{\rm RGO}}$ transmission curve quite closely resembles that of Sloan $u$ \citep{Doi2010}.

For UVEX, the sequence of observations at each pointing was $r$-$U_{\rm RGO}$-$g$. Before 2012 narrowband HeI 5856 exposures were also included but are not presented here. 
 The exposure time used in each of $U_{\rm RGO}$, $g$ and $r$ was 120, 30, and 30\,seconds, respectively. 
 For IPHAS the observing sequence was $H\alpha$-$r$-$i$.
 The $H\alpha$ filter exposure time was 120\,s throughout.  The majority of $i$ and $r$ frames were exposed for 10\,s and 30\,s respectively. There are two periods of exception to this: in the 2003 observing season, at survey start, the $r$ exposure time was 10\,s, while the $i$ exposure time was raised to 20\,s from 2010 October 29. 

The northern Galactic plane is covered via 7635 WFC fields that tessellate the footprint with, typically, a small overlap. In addition, each field is repeated with a shift of +5\,arcmin in RA and +5\,arcmin in Dec in order to fill in the gaps between the CCDs and also to minimize the effects of bad pixels and cosmic rays.  We refer to each pointing and its offset as a "field pair".  Quality checks were developed and applied to all the data, and those exposure sets ($r$, $i$ and $H\alpha$ -- or $U_{RGO}$, $g$ and $r$) rated as below standard were requeued for re-observation. The ID for each survey pointing is constructed using four digits,  starting by 0001 and rising with Right Ascension up to 7635, with an "o" straight after in the case of an offset pointing making up the field pair.

For a plot showing the footprint occupied by both surveys, the reader is referred to figure 2 presented by \cite{IPHASDR2}.  The difference now is that IPHAS observations are complete, filling the whole region between the boundaries at $-5^{\circ} < b < +5^{\circ}$, $30^{\circ} < \ell < 215^{\circ}$.  For UVEX, the coverage stops
just short near the celestial equator, at $RA = 110\fdg0$,
creating a triangular region of $\sim$33 sq.deg. (1.8\% of footprint) in which there is gradually reducing UVEX coverage of Galactic longitudes greater than $\ell \sim 205^{\circ}$.

\section{Data reduction and quality control}\label{sec:Red}

\subsection{Initial pipeline processing}\label{sec:pipe}

Over the 15 years of data taking, the observations passed from the INT to the Cambridge Astronomical Survey Unit (CASU) for processing.  A description of the pipeline and its conventions was given in the IPHAS DR2 paper \citep{IPHASDR2}.  Features specific to UVEX pipeline processing were noted by \cite{UVEX}.  For present purposes it is important to note that the pipeline (i) produces a photometric calibration based on nightly standards referred to a `run' mean, where a run is typically a period of a week or two of observing, (ii) places the astrometry onto the same reference frame as the 2MASS NIR survey.  In producing the IGAPS catalogue, a uniform calibration has been applied and the astrometry has been recomputed to place it in the Gaia DR2 frame \citep{2018A&A...616A...1G}. See details in sections \ref{sec:Cal} and \ref{sec:astro}, respectively.

\subsection{Quality control}
\label{sec:quality}

\begin{figure}\centering
 \resizebox{.8\hsize}{!}{\includegraphics{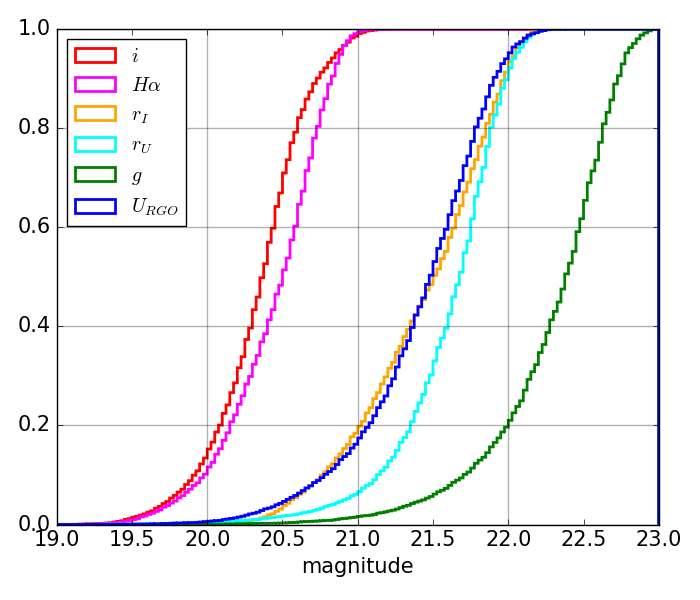}}
\resizebox{.8\hsize}{!}{\includegraphics{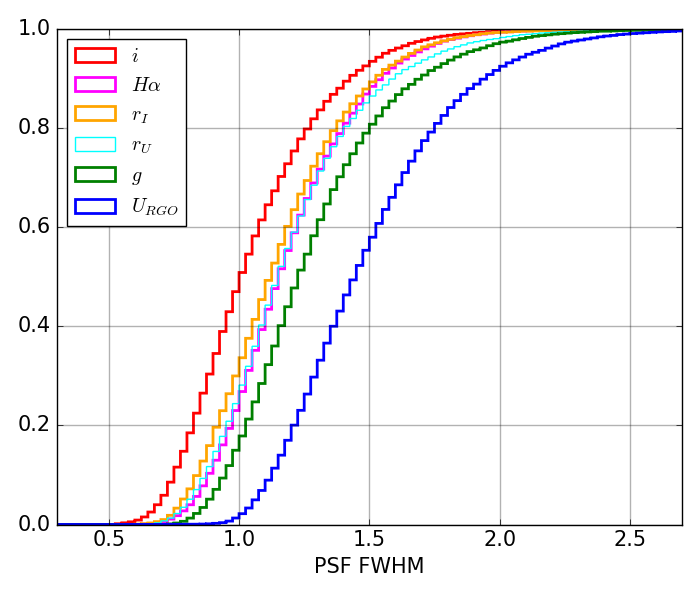}}
\caption{Top: Cumulative distribution of the $5\sigma$ limiting magnitude across all published survey fields for each of the five filters.  
Bottom: Cumulative distribution of the PSF FWHM for all fields included in the release, measured in the six filters. The PSF FWHM measures the effective image resolution that arises from the combination of atmospheric and dome seeing, and tracking accuracy.
}
\label{fig:5sig}
\end{figure}

Since the observations were collected over more than a decade using a common-user facility, a broad range in observing conditions necessarily exists within the survey databases. A variety of quality checks have been developed and applied to all fields as observed in both surveys. These checks were also used to assign a quality flag (or $fieldGrade$) from $A$ to $D$ to each field. See table \ref{tab:grades} for details on how this is implemented. The fields graded as $D$ were rejected and   the three filters re-observed when possible. In the absence of replacement, such fields were appraised individually and only kept if considered free of misleading artefacts. 
The different checks made are outlined below.

\begin{enumerate}
    \item Exposure Depth: In the top panel of figure~\ref{fig:5sig} we can see the $5\sigma$ magnitude limit distribution for all the exposure sets included in the data release. The limits are significantly better than 20 --for $r$ and $g$--, or 19 --for $i$ and $H\alpha$. The exposure sets that do not reach these limits are flagged as $fieldGrade=D$. We can see that some fields reach magnitude limits of 22 --in $r$--, 23 --in $g$--, and 21 --for $i$ and $H\alpha$.
    \item Ellipticity: The aim was that all included exposures would have mean ellipticity smaller than 0.3.  Exposures breaching this limit are labelled $fieldGrade=D$.
    Common values for the survey are in the range 0.15 to 0.20.
    \item Point spread function at full width half-maximum (PSF FWHM): Where possible, fields initially reported with PSF FWHM exceeding 2.5~arcsec were reobserved. 
    As can be seen in the lower panel of figure~\ref{fig:5sig}, the great majority of exposures return a PSF FWHM between 1 and 1.5\,arcsec in $r$.  And there is the expected trend that stellar images sharpen with increasing filter mean wavelength.
    \item Broad band scatter: Comparison with Pan-STARRS $r$, $g$, and $i$ data is central to the uniform calibration.  In making these comparisons, the standard deviation of individual-star photometric differences about the median offset ($stdps$) was computed. When this scatter in any one of the three filters exceeds 0.08, the IPHAS (or UVEX) $fieldGrade$ is set to $D$.  High scatter most likely indicates patchy cloud cover or gain problems.
    \item $H\alpha$ photometric scatter: Since the narrow band has no counterpart in Pan-STARRS, we use the photometric scatter computed between the $H\alpha$ exposures within a field pair to assess their quality. If the fraction of repeated stars exceeding pre-set thresholds in $|\Delta H\alpha|$ lies above the 98\% percentile in the distribution of all $H\alpha$ field pairs, both exposure sets involved are flagged as $fieldGrade=D$. Again, extreme behaviour most likely indicates patchy cloud cover or gain problems.
    \item Visual examination: Sets of images per field were individually reviewed by survey consortium members. A systematic by-eye examination of colour-magnitude and colour-colour diagrams was also carried out. When severe issues were reported, such as unexpectedly few stars, signs of patchy cloud cover, or pronounced read-out noise patterns, the exposure set would be rejected or given a $fieldGrade=D$ (if marginal and without an alternative).  
    \item Requirement for contemporaneous (3-filter) exposure sets: The survey strategies required the three IPHAS, or UVEX, filters at each pointing should be observed consecutively -- usually within an elapsed time of $\sim$5 min. All included exposure sets meet this criterion.
\end{enumerate}

\begin{figure*}\centering
 \resizebox{.38\hsize}{!}{\includegraphics{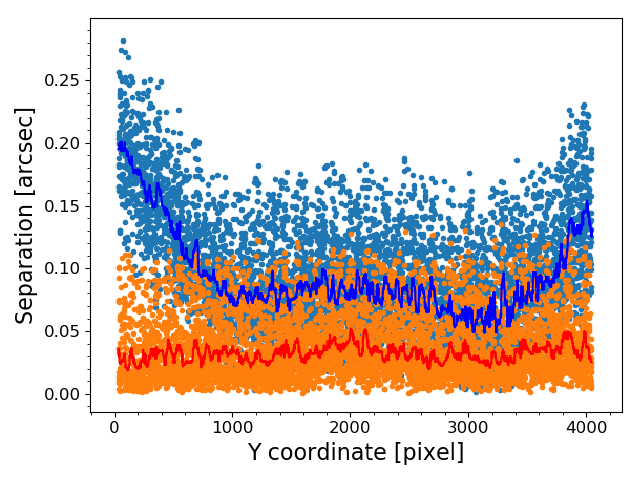}}
 \resizebox{.38\hsize}{!}{\includegraphics{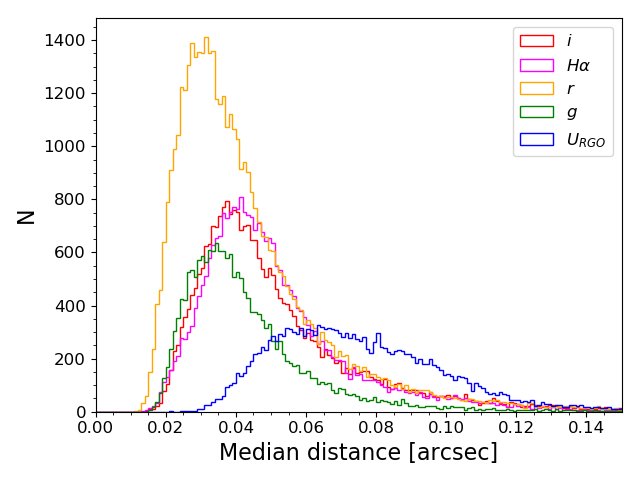}}
\caption{Left: Celestial position difference between the IGAPS catalogue and Gaia DR2 stars on CCD\#1 of INT image r908084. The original pipeline solution is shown in cyan and the refined solution in orange. The binned median curves are shown in blue and red, respectively.
Right: Histogram of the median celestial position difference for WFC CCD\#4 between IGAPS and Gaia DR2 by filter. The $r$ filter (orange) includes IPHAS and UVEX data.  The median differences are: 72 ($U_{\rm RGO}$), 39 ($g$), 38 ($r$), 46 ($H\alpha$) and 45 ($i$) mas.}
\label{fig:astdiff}
\end{figure*}

\section{Astrometry: resetting to the Gaia DR2 reference frame}\label{sec:astro}

The pipeline for the extraction of the survey data, as described in previous releases of IPHAS \citep{IPHASIDR,IPHASDR2}, sets the astrometric solutions using 2MASS \citep{2006AJ....131.1163S} as the reference. This was the best choice available at the start of survey observations.  Especially for very dense fields, source confusion can lead to a wrong world coordinate system (WCS) in the pipeline reduced images. Also for the blue bands in UVEX, particularly the $U_{\rm RGO}$ filter, the use of an IR survey as the astrometric reference can be problematic.

The natural choice for astrometric reference now is the Gaia DR2 \citep{2018A&A...616A...1G,2018A&A...616A...2L} reference frame.
The starting point for a refinement of the astrometry is the 2MASS-based per-CCD solution. The pipeline uses the zenithal polynomial projection \citep[ZPN, see][]{2002A&A...395.1077C} to map pixels to celestial coordinates. In this solution all even polynomial coefficients are set to 0, while the first order term (PV2\_1) is set to 1 and the third order term (PV2\_3) to 220. Occasionally, it was found that for the $U_{\rm RGO}$ filter a fifth order term (PV2\_5) also needed to be introduced. Free parameters in the solution were the elements of the CD matrix, which is used to transform pixel coordinates into projection plane coordinates, and the celestial coordinates of the reference point (CRVALn).

For the refinement of the astrometric solution using the Gaia DR2 catalogue we first remove IGAPS stars that are located close to the CCD border. We also remove very faint stars. The limit for removal is set as a threshold on the peak source height: the value chosen depends on the number of sources in the image, varying between 20 (in low stellar density fields) and 150 (high density fields) ADU. An exception is made for the $U_{\rm RGO}$ filter where the threshold is always 10 ADU.  Next, we search for Gaia DR2 sources within a 0.5 degree radius of the field centre. We then remove all sources that have a proper motion error in either Declination or Right Ascension greater than 3 mas/yr. The Gaia catalogue is then converted to the IGAPS observation epoch using the {\sc stilts} Gaia commands {\verb epochProp } and {\verb epochPropErr } \citep{2006ASPC..351..666T}.

The Gaia and the IGAPS catalogues are then matched using the match\_coordinates\_sky function in the {\sc astropy} package \citep{astropy:2013,astropy:2018}. Matches with a distance larger than 1.5 arcseconds are removed as spurious. Hence the initial astrometric solution of the pipeline needs to be better than this - which it usually is - if the search for a refined solution is going to succeed.  In the rare cases where the pipeline solution is worse than this, a good enough initial astrometric solution needs to be found by hand.

As the ZPN projection cannot be inverted, its coefficients need to be found iteratively. We are using the {\sc Python} package {\sc lmfit} \citep{newville_matthew_2014_11813} with the default Levenberg Marquart algorithm for finding the iterative solution. The fitting function converts the IGAPS pixel positions into celestial coordinates using the ZPN parameters and calculates the separation to the matched Gaia source, which is minimized.
As the solution depends on the initial parameters, we run the algorithm with 10 different starting parameter sets: the original pipeline solution; the set of median coefficients for the CDn\_m and PV2\_3 values of the filter; plus 8 sets where CRPIXn, CDn\_m and PV2\_3 are randomly adjusted by up to 5\% from the original pipeline solution values.  For the $U_{\rm RGO}$ filter PV2\_5 is treated in the same way as PV2\_3.

The best solution among the 10 tries is found as follows.  The separation in arcseconds between IGAPS and Gaia is binned with bin sizes of up to 51 stars, depending on the number of stars on the CCD, along the longer axis of the CCD and the median in each bin is calculated (solid blue and red lines in figure \ref{fig:astdiff}). The solution that has the lowest maximum bin celestial position difference is kept as the best astrometric solution. The median of all bins is kept as the astrometric error to be reported in the final IGAPS catalogue (column $posErr$). 

The left hand panel of figure~\ref{fig:astdiff} shows an example of an initial pipeline and a final astrometric solution. The {\em maximum bin} celestial position difference relative to the Gaia frame in $r$ -- the filter that provides the position for the great majority of sources in the final catalogue -- in this example was reduced from 0.23 arcsec initially to 0.061 arcsec in the refined solution.  The right hand panel of figure~\ref{fig:astdiff} shows that the performance in CCD~4, where the optical axis of the camera falls, is generally to achieve median position differences under 0.1\,arcsec. It also illustrates the point that the solutions for $U_{\rm RGO}$ are least tight.  Experiments with the data suggest the main contribution to the error budget is due to the optical properties of the $U_{\rm RGO}$ filter as a liquid filter, with differential chromatic refraction playing only a minor role. However the improvement this represents for $U_{\rm RGO}$ is arguably greater than for the other filters, in that the original astrometry was often so poor that cross-matching of this filter to the others would fail for much of the camera footprint.  In this respect, a recalculation of the astrometry was a pre-requisite for the successful construction of the IGAPS catalogue.

\section{Global photometric calibration} \label{sec:Cal}

The approach to global calibration is as follows.  Since the entire IGAPS footprint falls within that of the Pan-STARRS survey \citep{Chambers16}, we have chosen to tie IGAPS $g$, $r$, and $i$ -- the photometric bands in common -- to the Pan-STARRS scale \citep{2016arXiv161205242M}.  By doing this it is possible to piggy-back on the high quality 'Ubercal' that benefitted particularly from the much larger 3-sq.deg. field of view of the Pan-STARRS instrument \citep{Magnier2013}.  With the $g$, $r$, $i$ calibration in place, we are then able to link in the narrowband $H\alpha$, as described below.  A global calibration of $U_{RGO}$ is not attempted at this time (see Section~\ref{sec:URGO} for more comment).

Previous IPHAS data releases have provided photometry adopting the Vega zero-magnitude scale.  We continue to do this here, whilst also offering the option in the catalogue of magnitudes in the (Pan-STARRS) AB system.

\subsection{Calibration of $g$, $r$ and $i$, with respect to Pan-STARRS}\label{ssec:gri}

\begin{figure}\centering
 \resizebox{.48\hsize}{!}{\includegraphics{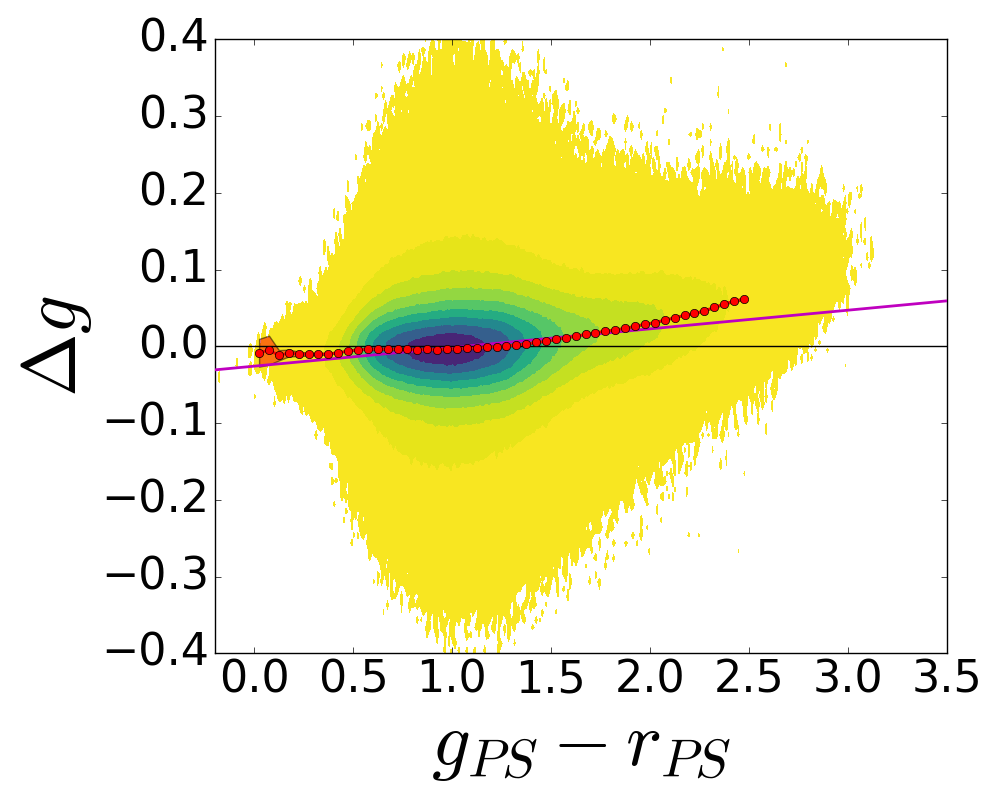}}
\resizebox{.48\hsize}{!}{\includegraphics{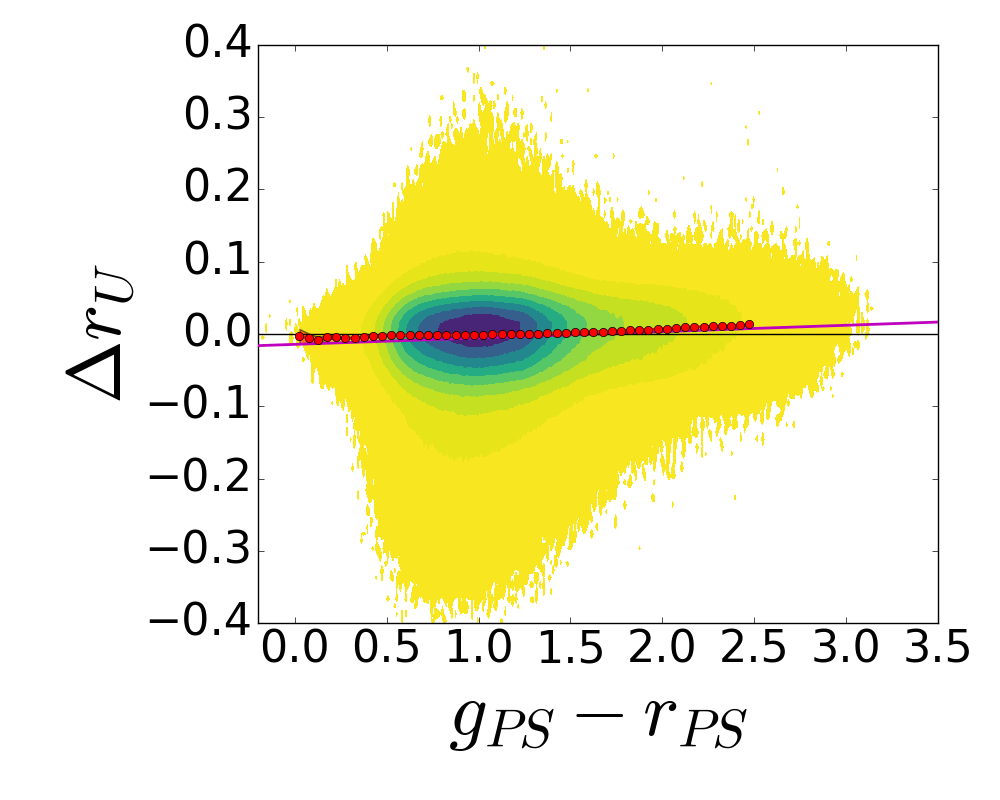}}
\resizebox{.48\hsize}{!}{\includegraphics{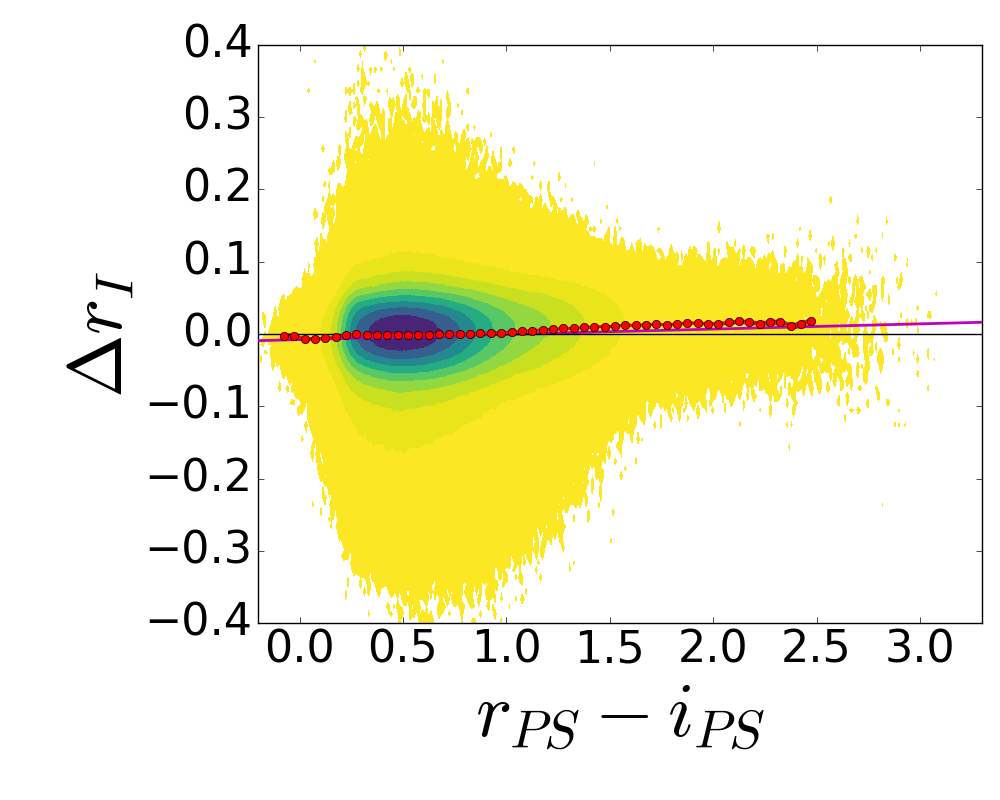}}
\resizebox{.48\hsize}{!}{\includegraphics{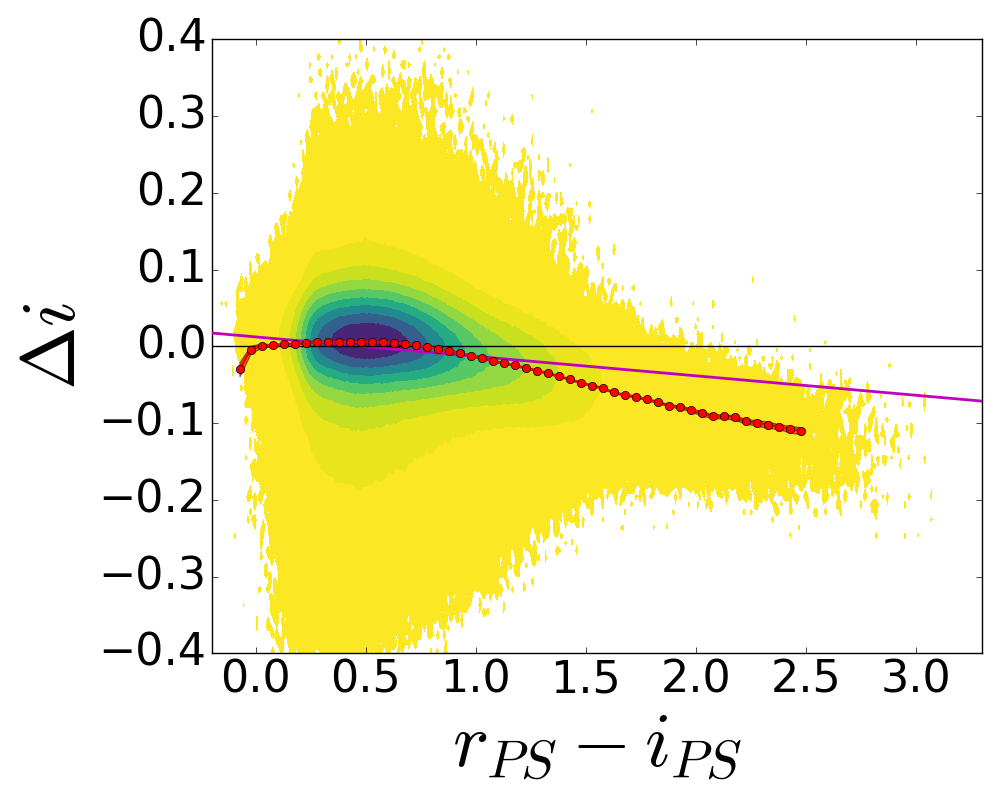}}
 \caption{ Differences between IGAPS and Pan-STARRS magnitudes -after taking out the raw per CCD median shift- vs Pan-STARRS colour. Data from the range $50^{\circ}< \ell <70^{\circ}$,  $-5^{\circ}< b <+5^{\circ}$ are shown. Top-left: $\Delta g$ vs ($g-r$), top-right: $\Delta r_{U}$ vs ($g-r$), bottom-left: $\Delta r_{I}$ vs ($r-i$), bottom-right: $\Delta i$ vs ($r-i$). Only stars with $14<g_{ps}<20$, $13<r_{ps}<19$ or $12.5<i_{ps}<18.5$ are used in these plots. The magenta line is the fitting line. The red dots follow the running median for each 0.05\,mag bin showing where the trends deviate.  The false colour scale indicates density of sources in each bin on a square root scale with yellow representing the lowest density of at least 4 sources per 0.02x0.02\,mag$^2$ bin.}
\label{Fig:cterm}
\end{figure}

The calibration was carried out on a chip by chip basis, computing the median differences between IGAPS and Pan-STARRS magnitudes in each of the three filters, after allowing for a colour term as needed.
In order to compute these, we plotted 
the differences in magnitude as a function of colour, paying attention to sky location. 
Specifically, we computed the shift gradient as a function of colour for a set of 
boxes spanning the survey footprint. No significant trend was apparent in any filter, although variation in the gradient by up to $\pm$0.01 was noted.  We provide an example of the colour behaviour for each of the filters in figure \ref{Fig:cterm}.
 We concluded that, overall, there is no need for a colour term in handling the $r$ band, whilst correction is appropriate for $g$ and $i$.

The final calibration shifts applied per band 
per CCD are:
\begin{equation}
    \begin{aligned}
\Delta ZP_r &= {\rm median}(r^p +0.125-r_{PS}) \\
\Delta ZP_g &= {\rm median}\left[ (g^p-0.110-g_{PS})-0.040\cdot(g_{PS}-r_{PS}) \right]\\ 
\Delta ZP_i &= {\rm median}\left[ (i^p+0.368-i_{PS})+0.060\cdot(r_{PS}-i_{PS})\right]
    \end{aligned}
\end{equation}
where the superscript $p$ indicates the Vega magnitudes from the pipeline and the constants in the first right-hand-side brackets are the transformation coefficients from Vega to AB magnitudes in the INT filter system. To assure the quality of the shift calculation, only those stars within a specified magnitude range were taken into account, in order to avoid bright stars subject to saturation, and fainter objects with relatively noisy magnitudes.
The ranges used were $15<g<19$, $14.5<r<18.5$, and $13.5<i<17.5$~mag.

\begin{figure*}\centering
\resizebox{.99\hsize}{!}{\includegraphics{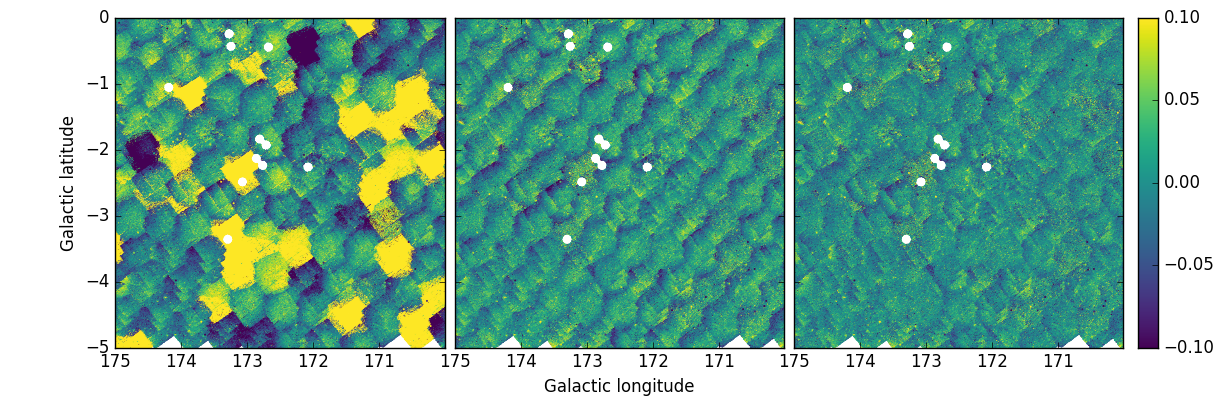}}
 \caption{5x5deg$^2$ box at 170$\degr$<$l$<175$\degr$, -5$\degr$<$b$<0$\degr$ as observed in UVEX $r$ band. This region is picked as representative of the more difficult and changeable (winter) observing conditions.  Colours shown indicate the magnitude differences with respect to Pan-STARRS. Left: The differences before calibration; centre: differences after calibration; right: differences after the small additional illumination correction is applied. White holes are excluded regions around bright stars.
 }
\label{Fig:difps}
\end{figure*}

Once the shift for each CCD and filter is computed, the calibrated AB magnitude for each star is recovered.  This proceeds by first calculating the corrected $r$ magnitude in the AB system, via:
\begin{equation}
    r_{AB} = r^p + 0.125 - \Delta ZP_r
\end{equation}    
The ground is then prepared for finding the $g_{AB}$ and $i_{AB}$ magnitudes taking into account the relevant colour term:
\begin{equation}
    \begin{aligned}
    g_{AB} &= \frac{1}{1.040}\cdot\left[g^p - 0.110 - \Delta ZP_g + 0.040 \cdot r_{AB}\right]\\
    i_{AB} &= \frac{1}{1.060}\cdot\left[i^p + 0.368 - \Delta ZP_i + 0.060\cdot r_{AB}\right]
    \end{aligned}
\end{equation}
Then, finally, the Vega corrected magnitudes are computed from these AB alternates 
using the shifts appropriate in the Pan-STARRS filter system:
\begin{equation}
    \begin{aligned}
    r &= r_{AB} - 0.121\\
    g &= g_{AB} + 0.110\\
    i &= i_{AB} - 0.344
    \end{aligned}
\end{equation}
An example of this calibration step operating in one 5x5\,sq.deg. box is shown in the first two panels of figure \ref{Fig:difps}.

For faint red stars, when an $i$ magnitude is available but not $r$, the final $i$ magnitude is computed without taking into account the colour term. In such a case, the photometric error is raised to acknowledge this by adding in 0.05 mag, in quadrature. The same remedy is adopted for the much rarer instances of blue/faint objects for which $g$ is available but not $r$.

The standard deviation of the differences relative to Pan-STARRS for each CCD chip ($stdps$), computed alongside the median shift (equation 1) is retained to serve as a measure of the quality of the IGAPS photometry. For example, a photometric gradient across a chip, due to cloud or a focus change, will not be removed by the calibration shift, but will increase the recorded standard deviation. This datum is used within the seaming process in deciding which detections to identify as primary in the final source catalogue.

\subsection{Final adjustment of the illumination correction}
\label{sec:ilu}

It is a part of the pipeline extraction to compute and apply seasonally-adjusted illumination corrections to all survey data.  Whilst this does most of the job, some residual unevenness became apparent in assembling the data for this first merged catalogue. Specifically, in the second column of figure~\ref{Fig:difps} a subtle diagonal rippling pattern due to slowly varying 'illumination' can be noticed that is systematic with position within the CCD mosaic. To deal with this we make a further global adjustment in the style of an illumination correction in order to reduce the ripple.  

\begin{figure}\centering
 \resizebox{.85\hsize}{!}{\includegraphics{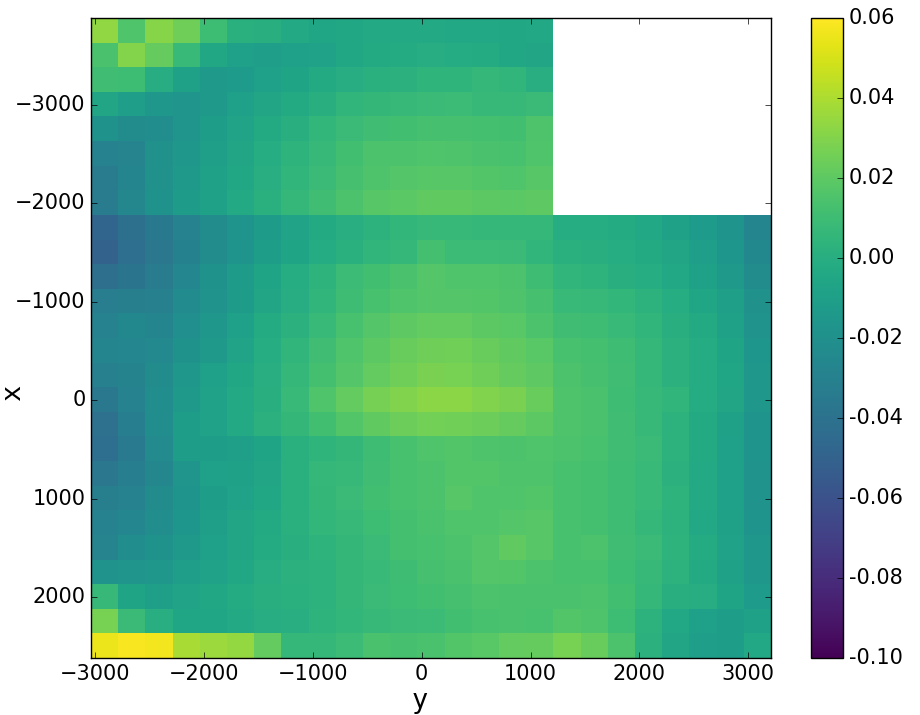}}
\caption{Differences between Pan-STARRS and IGAPS $i$ band magnitudes in pixel space within the 4-CCD mosaic. Median values are plotted for each 250x250 pixel$^2$ bin. The numerically strongest deviation is in the y coordinate (bluer colours to left and right in the figure).
}
\label{fig:ilcor}
\end{figure}

To analyse this effect in more detail we examined the differences in magnitude between our survey and Pan-STARRS as a function of position within the CCD mosaic. Summing the observations and computing the median value for each 250x250 pixel$^2$ bin, we obtain plots like the $i$ band example shown in figure~\ref{fig:ilcor}. In this exercise, we have used only high quality stars (errBits=0, see section~\ref{sec:flags}). For the $g$ band, this takes out of consideration stars affected by a blemish on the filter (see section~\ref{sec:gband}).

\begin{figure*}\centering
 \resizebox{.98\hsize}{!}{\includegraphics{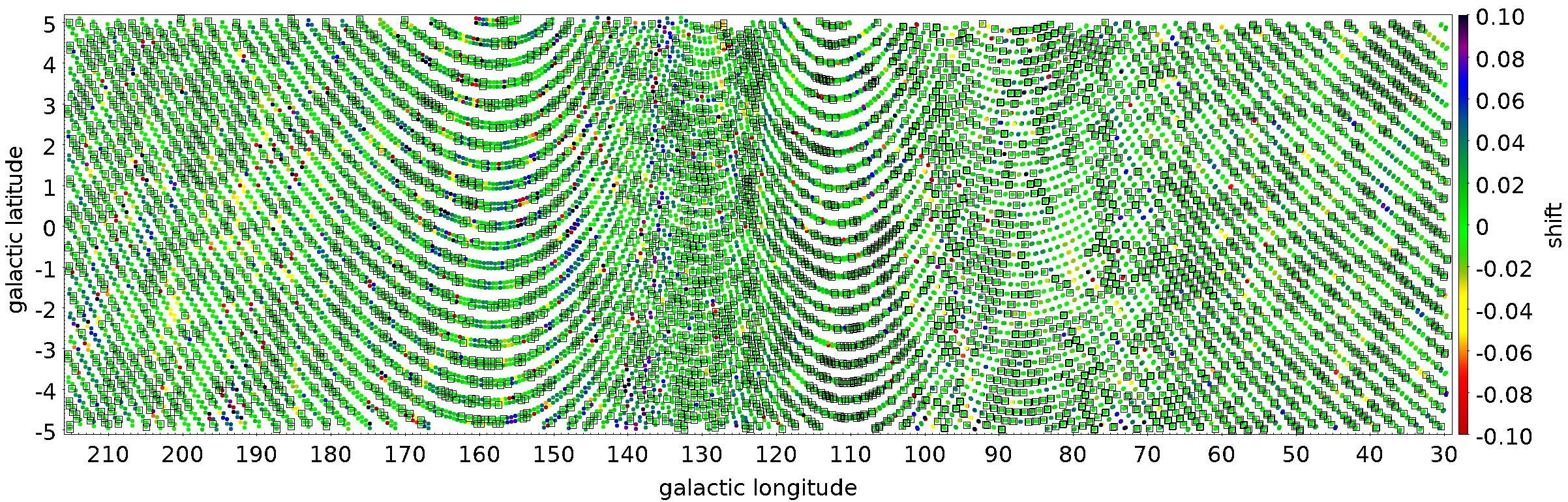}}
\caption{To illustrate the outcome of the $H\alpha$ calibration, the Galactic Plane footprint is shown with all the fields marked as points. Colour indicates the shift applied to the $H\alpha$ zeropoint, according to the Glazebrook correction, while the black squares indicate the fields used as anchors.}
\label{fig:glaz}
\end{figure*}

In all filters, we found a remnant pattern at the level of a few hundredths of a magnitude that can be partially modelled out. We tried a range of fit options, including both a radial pattern and a double parabola in the $x$ and $y$ pixel coordinate in the image plane, and found that the smallest residuals were associated with fitting a parabola in only the $y$ pixel (i.e. Right Ascension) direction. This result was also found by \cite{2013A&A...549A..78M}, although these authors did not have enough measurements to obtain a statistically significant outcome that warranted application to the data.  We separately fit the correction for the four different filters $i$, $r_I$, $r_U$ and $g$, and concluded that the resultant curves 
are sufficiently alike that there is no compelling need to retain and apply them independently.  Hence a single correction curve was constructed combining all $g$, $r$, $i$ magnitudes and was applied uniformly to all bands, including $H\alpha$ and $U_{RGO}$. 
This approach means that there is no effect on calibrated stellar colours in the catalogue.  
The functional form of this correction is:
\begin{equation}
\begin{aligned}
\mathrm{Dmag} = -4.93\times10^{-9} y^2 +4.35\times10^{-6} y + 0.014
\end{aligned}
\end{equation}
where Dmag is the correction in magnitudes to be applied, 
and $y$ is the pixel coordinate within the field, with origin at the optical axis.  This fit to the data has an error ($\sigma$) of 0.008\,mag.
The result of this correction can be seen in the right panel of figure~\ref{Fig:difps}, where the ripples are damped down.

\begin{figure}\centering
 \resizebox{.85\hsize}{!}{\includegraphics{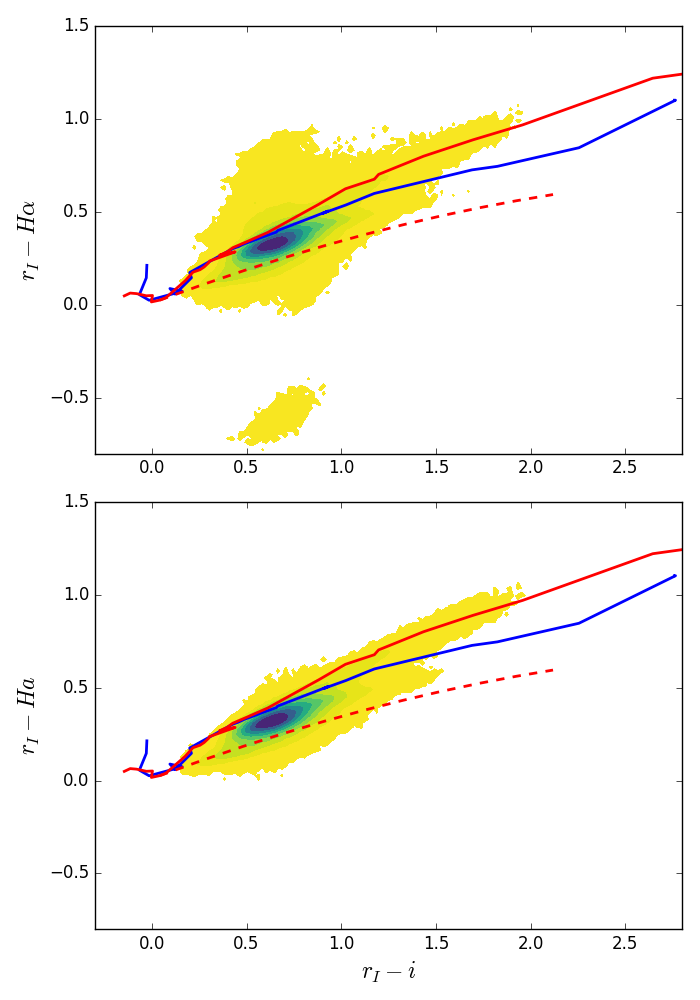}}
\caption{$r-H\alpha$ vs $r-i$ diagrams for the region $165^{\circ}<l<170^{\circ}$ before (top) and after (bottom) the Glazebrook calibration, using sources with $r_I<19$ and errBits=0. Lines indicate the expected sequences: unreddened main sequence in red, giants sequence in blue, and reddening line  for an A2V star up to $A_V$=10 in dashed red. See Appendix D for photometric colour tables. As explained by \cite{IPHAS} and \cite{Sale09}, the elongation of the main stellar locus is due to the combined effects of interstellar extinction and intrinsic colour. The false colour scale indicates density of sources in each bin on a square root scale with yellow representing the lowest density of at least 4 sources per 0.02x0.02\,mag$^2$ bin.}
\label{fig:cc}
\end{figure}

\subsection{$H\alpha$ calibration}

Narrowband $H\alpha$ is the signature filter of IPHAS that cannot be calibrated against other wide-field surveys. Accordingly an internal method is needed.  

Since this band is embedded within the $r$ band, it means that the same calibration shifts applied to the $r$ band can also be applied to the $H\alpha$ band of each observing sequence as a first estimate.  Indeed the extraction pipeline assumes that in stable weather there should be a constant offset between the $r$ and $H\alpha$ zeropoints, and it is applied as a matter of routine. This offset was taken to be 3.14 at the time of IPHAS DR2 \citep{IPHASDR2}.  Based on more recent data, we now regard 3.115 as the better value.  This update has been obtained by folding the spectrum of Vega  \citep[CALSPEC stis\_009,][]{2014PASP..126..711B} with the ING measured filter curves and an atmosphere calculated with ESO SkyCalc \citep{2012A&A...543A..92N, 2013A&A...560A..91J} for La Silla (similar altitude to La Palma), an airmass of 1.2 \citep[as used by Pan-STARRS,][and close to our survey median of 1.15]{2012ApJ...750...99T} and a precipitable water vapour (PWV) content of 5\,mm \citep{2009SPIE.7475E..1HG}. Optical surfaces were not taken into account, as precise measurements of them were not available and they are expected to be grey over the r-band filter. 

The different versions of the CALSPEC Vega spectrum introduce only a small change of 0.003 mag in the offset calculation. A similar change can be achieved by using different measurements of the filter curves obtained over the survey years at the ING. The effect of airmass is a lowering of about 0.003 mag per 0.1 airmass change over the range that survey observations were taken. Finally an increase of about 0.0025 mag per 5 mm PWV is found. The $H\alpha$ zeropoint offset from r found this way is 3.137.
However, when comparing synthetic stellar locations to the data it was found that the location of A0 dwarf stars is not at zero colour in r-$H\alpha$, as would be expected by definition in the Vega system. The cause of this lies in a unique aspect of the standard star Vega, namely it being a fast rotator viewed nearly pole on \citep{2010ApJ...712..250H}, which introduces a difference in the $H\alpha$ line profile when compared to other A0 dwarf stars. Indeed, the other CALSPEC A0V standards, HD116405 and HD180609, show a lower value of the zero point offset between r and $H\alpha$. The value we used for the offset is the average of the offsets derived from these two stars with the different filter profiles measured over the years. Using this value of 3.115 for the zeropoint offset between r and $H\alpha$ is equivalent to saying that the magnitude of Vega in the $H\alpha$ filter is 0.022 mag fainter than in the broadband filters.

\begin{table*}\centering
\begin{tabular}{ccccccc}
\hline
Filter  & $\langle f_\lambda \rangle$   & EW    & $\lambda_0$ & $\lambda_p$ & \multicolumn{2}{c}{Vega magnitude}   \\
  & (erg cm$^{-2}$ s$^{-1}$) \AA$^{-1}$ & (\AA) & (\AA) & (\AA) & AB & Vega system \\
\hline
$U_{RGO}$ &  $4.24 \times 10^{-9}$ & 138.8 & 3646 & 3640 & 0.742 & 0.023 \\
g         &  $4.98 \times 10^{-9}$ & 716.7 & 4874 & 4860 & $-$0.088 & 0.023 \\
r         &  $2.44 \times 10^{-9}$ & 745.3 & 6224 & 6212 & 0.153 & 0.023 \\
$H\alpha$ &  $1.79 \times 10^{-9}$ &  57.1 & 6568 & 6568 & 0.373 & 0.045 \\
i         &  $1.29 \times 10^{-9}$ & 708.2 & 7677 & 7664 & 0.393 & 0.023 \\
\hline
\end{tabular}
\caption{As table 2 in \cite{IPHASDR2} with values for all the survey filters. Mean monochromatic flux of Vega, filter equivalent width, mean photon and pivot wavelengths as defined in \cite{2012PASP..124..140B} are given, along with the calculated AB and defined Vega system magnitudes for the CALSPEC Vega spectrum stis\_009 \citep[the Vega broad band magnitude is from][]{Bohlin2007}. Note that the catalogue data for $U_{RGO}$ is not globally calibrated and the broad band filters $g$, $r$ and $i$ are transformed onto the Pan-STARRS photometric system.}\label{tab:flux}
\end{table*}

To deal with random shifts due to poor and variable weather, a second correction is applied that seeks to minimize the differences between $H\alpha$ magnitudes --after illumination correction is applied -- in the zones of overlap between fields \citep{1994MNRAS.266...65G}. This requires the selection of the best fields, or anchors, that are fixed under the assumption their photometry needs no further correction. 
The fields to be used as anchors are carefully selected taking into account: the standard deviation of the magnitude differences with Pan-STARRS ($stdps$) in $r$ and $i$, to avoid magnitude gradients in the field; the number of stars crossmatched with the Pan-STARRS catalogue, to ensure adequate statistics; the median value of the magnitude differences between the field and its offset pair, taken to indicate a stable night. As a final precaution, the $(H\alpha-r)$ vs. $(r-i)$ diagram for each potential anchor field was inspected to check for consistent placement of the main stellar locus. The shifts applied and the selection of anchors can be seen in figure~\ref{fig:glaz}. In figure~\ref{fig:cc} we can see the $r-i$ vs $r-H\alpha$ diagrams for the region $165\degr<l<170\degr$ before and after the Glazebrook calibration -- the improvement is clear.

\section{Artefact mitigation}
\label{sec:artefacts}

With astrometry re-aligned to the Gaia DR2 reference frame, and a uniform calibration in place, the next steps are to conduct some final cleaning and flagging.

\subsection{Mitigation of satellite trails and other linear artefacts}
\label{sec:trails}

The night sky is criss-crossed by satellites and meteors liable to leave bright trails in exposed survey images, essentially at random.  It is far and away most common that the photometry of any given detected object is adversely affected in one band only by this unwanted extra light.  Nevertheless, it is important to the value of the final merged catalogue that instances of the problem are brought to the user's attention.

To achieve this, we have visually inspected composite plots of IPHAS-bands and UVEX-band catalogued objects, noting instances of trails and other linear artefacts. The affected individual-filter flux tables are then visited in order to mark and flag these features. Satellite trails usually show up very easily in these plots. But, in more ambiguous cases, the images themselves have also been checked. Strips of width 30 pixels -- have been computed and placed on all noted linear streaks, and have been used to flag all sources falling within them as at risk.
This intensive visual inspection also brought to the fore other linear structures such as spikes due to bright stars, crosstalk, and read-out problems and meant they too could be flagged.

\subsection{Masking of localised PSF distortion on the $g$-band filter}
\label{sec:gband}

With the accumulation of more and more survey data and the release of Pan-STARRS data \citep{Chambers16}, it became possible to co-add large numbers of detected-source magnitude offsets referred to pixel position in the image plane. This reveals any localised variations in photometric performance that might otherwise be missed.  In the case of the $g$ band, this procedure revealed a clear distortion towards the edge of the image plane compromising the extracted photometry.  Subsequent visual inspection of the filter by observatory staff confirmed the presence of a blemish near its edge, in the right place to be linked with the evident photometric distortion. 

Since flat field frames taken through the $g$ filter did not betray the problem, a transmission change could not be implicated. Instead a change in the character of the point-spread function (PSF) had to be involved.  Further checking revealed that point-source morphologies returned by the extraction pipeline were changing (sharpening) in the region of the blemish.  Since the PSF and associated aperture corrections are computed in the pipeline per CCD, these changes over the smaller area of distortion would not be tracked adequately and would lead to over-large aperture corrections in the affected area of the chip. 

After mapping the regions affected (and the variations as a function of date of observation, due to rotation of the filter within its holder from time to time), we are able to flag the stars falling in them.  This is done at two levels of impact. We have defined as the inner, most severely affected region within the camera footprint, those locations where the photometric discrepancy exceeds 0.1 mag, while the threshold set for the outer region is 0.05 mag.  The lower of these thresholds corresponds to roughly 4 times the median shift elsewhere in the footprint (computed for stars in the range $15 < g < 19$).  The g-band detections masked in this way always fall near the edge of the imaged area, within an area amounting to roughly 0.07 of the total.  In terms of primary detections listed in the catalogue, the choices made in the seaming algorithm bring the g-mask flagged fraction down to 0.015.  
More detail on how the $g$ mask is imposed is given in supplementary materials (Appendix~\ref{sec:gmask_supp}).

\begin{table*}\centering
\begin{tabular}{lccrrcl}
\hline
Star    & RA & DEC& \multicolumn{1}{c}{$\ell$} & \multicolumn{1}{c}{b} & $V$ & IDs of affected fields    \\\hline
Capella &  05 16 41.36& +45 59 52.77&  162.589 & +4.566 & 0.08 &2298, 2298o\\
Deneb &   20 41 25.92& +45 16 49.22&   84.285 & +1.998 & 1.25&6116, 6116o, 6083o,6093\\
Elnath&  05 26 17.51 &+28 36 26.83& 177.994 & -3.745 & 1.65 & 2416,2416o,2452,2452o\\
Alhena& 06 37 42.71& +16 23 57.41 & 196.774 & +4.453 & 1.92 & 3720,3720o,3690,3690o\\
$\gamma$ Cyg&  20 22 13.70&+40 15 24.04 & 78.149 & 1.867 & 2.23 & 5868,5868o,5831,5831o,5855,5855o\\
$\beta$ Cas & 00 09 10.69 & +59 08 59.21 & 117.528 & -3.278 & 2.27&0043,0043o,0052,0052o,0066\\
$\gamma$ Cas& 00 56 42.53& +60 43 00.27 & 123.577 & -2.148 & 2.39& 0324,0302,0302o,0296,0296o\\
$\delta$ Cas &  01 25 48.95& +60 14 07.02 & 127.190 & -2.352 & 2.68&0459o,0475o,0477,0477o\\
$\mu$ Gem &  06 22 57.63& +22 30 48.90 & 189.727 & 4.169 & 2.87&3413,3413o,3428\\
$\gamma$ Per &  03 04 47.79& +53 30 23.17 & 142.067 & -4.337 & 2.93 & 1051o,1055,1055o\\
$\zeta$ Aql& 19 05 24.61& +13 51 48.52 & 46.854 & +3.245 & 2.99 & 4483,4483o\\
$\epsilon$ Aur &  05 01 58.13& +43 49 23.87 & 162.788 & +1.179 & 2.99 & 2084,2084o,2106,2106o,2119\\
\hline
\end{tabular}
\caption{Stars brighter than $V = 3$ mag. located within the IGAPS footprint. It is recommended that catalogue users seeking photometry in the vicinity of these objects, especially, should check images  (Greimel et al, in prep.) to better understand the likely impact they have on the photometry.  For convenience both celestial and Galactic coordinates are given.}
\label{tab:brightest}
\end{table*}

\subsection{Bright stars, ghosts and read-out problems}
\label{sec:other-artefacts}

Bright stars can affect the photometry of other stars nearby. Not only that, but features in e.g. the diffraction spikes are sometimes picked up as sources by the pipeline. To support screening these out, we have identified all the stars in the Bright Star Catalogue \citep{Hoffleit1991} that are brighter than $V = 7$ in the survey area and have flagged all catalogued sources that lie within a radius of 5 arcmin of any of them. For sources brighter than 4th magnitude, this radius is raised to 10 arcmin.  Clearly some real sources that happen to be close to bright stars will be caught up in this, and flagged.  Interested users of the catalogue are encouraged to check the images (when available) in these instances, remembering that the background level is higher in these flagged regions with the result that sources in them may not be as well background-subtracted as sources in the wider unaffected field.  

Bright stars outside the field can also create spikes due to reflections in the telescope optics. When linear, these will have been flagged as part of the procedure described above in section~\ref{sec:trails}.  But occasional, more complex structures are likely to be missed.  In this category we place the structured dominantly-circular ghosts of stars brighter than $V = 4$.  These are obvious in the processed images and also show up as rings in wider-area plots of catalogued objects.

As the Wide Field Camera aged during the execution of IPHAS and UVEX, electronic glitches during read-out -- creating jumps in the background level -- became progressively more frequent. In cases where the whole image is affected by tell-tale strips and lines, it is discarded. But sometimes this issue affects just a small portion of one CCD. In cases like this, the image is retained if there is no alternative exposure available, while the sources in the minority problematic regions are flagged.

\subsection{Saturation level and the brightest stars}

Stellar images typically begin to saturate at magnitudes between 12 and 13.  Catalogued objects affected by this are flagged.  The precise saturation magnitude in an exposure is somewhat dependent on the seeing and sky conditions, both of which varied significantly over the 15 years of data gathering. 

It is worth noting that there are some extremely bright stars in the footprint that not only saturate but have a major detrimental effect on the photometry collected from the whole CCD in which they are imaged, and more.  In the most extreme case of Capella, nearly the entire 4-CCD mosaic is compromised. Such objects create rings, bright spikes and halos, ghosting between CCDs, as already mentioned in Section~\ref{sec:other-artefacts}. In table \ref{tab:brightest} we list the stars brighter than $V=$3\,mag in the footprint that are most challenging in this regard.

\section{Generation of the source catalogue}
\label{sec:cat}

\subsection{Catalogue naming conventions and warning flags}\label{sec:flags}
The detailed description of columns in the catalogue will be given in Appendix \ref{sec:AppTables}. Here we explain the meaning for some of the columns.

The name for each source, as suggested by the IAU,  is uniquely identified by
an IAU-style designation of the form 'IGAPS JHHMMSS.ss+DDMMSS.s', where the name of the catalogue IGAPS is omitted in the catalogue. 
 The coordinates of the source are also present in decimal degrees and in Galactic coordinates in columns $RA$, $DEC$, $gal\_long$, and $gal\_lat$. The coordinates come with an error ($posErr$) computed as indicated in Sect.\,\ref{sec:astro}. Since each source can be measured in up to six different bands, we always use as a reference $r_I$ if available. If it is not, then we will use, in order of preference, the coordinates extracted from the following bands:  $r_U$, $i$, $H\alpha$, and $g$. The differences in astrometry between the designated coordinates and the individual band coordinates can be found in $mDeltaRA$ and $mDeltaDec$ for each of the filters --except for $r_I$, that it is not included since being the primary source for the astrometry, it would always be zero if available. We provide another identifier for each band in $mdetectionID$, created by adding the run number of the original image, the ccd number and the detection number within this ccd, i.e. '\#run-\#ccd-\#detection'. A general $sourceID$ is chosen from those, using the same priority as for the coordinates, i.e. $r_I$, $r_U$, $i$, $H\alpha$, and $g$.

For each band we have a flag ($mClass$) indicating whether a source looks like a star ($mClass=-1$), an extended object ($mClass=1$) or noise ($mClass=0$). It can also indicate a probable star ($mClass=-2$) or a probable extended object ($mClass=-3$). A general $mergedClass$ flag will be set up to the same values if the different $mClass$ for all the available bands agree. Otherwise it will be set to 99. From the combination of these classes, we compute the probability for a source to be a star, noise or an extended source ($pStar$, $pNoise$, $pGalaxy$).

Boolean flags are also set up indicating whether the source in a given band is affected by deblending, saturation, vignetting, trails, truncation for being close to the edge of the ccd, or if it is close to a bad pixel. For each source and band, the user can also find the ellipticity, the median Julian date of the observation, and the seeing.

As a summary of the information provided by different bands, some final boolean flags are also available: $brightNeighb$ if the object is 
 located within a radius of 5 arcmin from an source brighter than V=7 according to the Bright Star Catalogue \citep{Hoffleit1991}, or within 10 arcmin if the neighbour is brighter than V = 4, $deblend$ if there is another source nearby, and $saturated$ if it saturates in one of the bands.  $nBands$ indicates the number of bands available for each source from the six possible $i$, $H\alpha$, $r_I$, $r_U$, $g$, $U_{RGO}$. $nObs_I$ is the number of IPHAS repeat observations available for this source and $nObs_U$ is the same for UVEX.

Another global quality measure provided is $errBits$.  It will be the addition of: 1 if the source has a bright neighbour; 2 if it is a deblend with another source in any band; 4 if it has been flagged as next to a trail in any band; 8 if it is saturated in any band; 16 if it is in the outer masking of the $g$ band blemish; 64 if the source is vignetted near the corner of CCD 3 in any band; 128 if it is in the inner mask of the $g$ band blemish; 256 if it is truncated near the CCD border in any band; 32768 if the source sits on a bad pixel (in any band). If ErrBits is not equal to 0, the user should exercise care when using the information provided for the source.

\subsection{Bandmerging and primary detection selection}
\label{sec:merging}

The merging of the different bands involves two steps. First, the three contemporaneous bands for each of IPHAS and UVEX are merged. We use the \textsc{tmatch} tool within {\sc stilts}  \citep{2006ASPC..351..666T} to obtain tables collecting together information on the three bands for each source, adopting an upper limit on the on-sky crossmatch radius of 1 arcsec.  With the re-working of the astrometry into the Gaia DR2 reference frame, it might seem that a tighter limit of e.g. 0.5 arcsec could be applied.  Whilst this is almost always true (see Section \ref{sec:difastro}), we used the more forgiving 1 arcsec bound to allow for the optical differences internal to the separate filter sets of IPHAS (including $H\alpha$ narrowband) and UVEX (including $U_{RGO}$).  It also gives more room to keep high proper motion counterparts together on merging the IPHAS and UVEX $r$ observations.  Sources missing a detection in one or more filters are retained in this process, with the columns for the missing band(s) left empty.  

Before the final UVEX-IPHAS merging, we must take into account the normal situation that a source in either catalogue has typically been detected in a given band more than once.  This arises from the standard observing pattern of obtaining a pair of offset exposures for every filter and field (a practice aimed at eliminating as far as possible the on-sky gaps that would otherwise exist due to the WFC's inter-CCD gaps). To bring to the fore the best available data, we do not stack information from repeat measures, but instead select the best measurement per source. To do that, we prioritise according to the following rules. If there is no clear winner at any one step, we then move to the next:
\begin{enumerate}
    \item Choose detection with greater number of bands available.
    \item Reject $fieldGrade$=D if other options are available.
    \item Choose detection with smallest $errBits$.
    \item Pick the detection with the smaller photometric dispersion in the Pan-STARRS comparison, using the $stdps$ flag. 
    \item Choose best seeing. 
    \item Select detection closest to the optical axis of the exposure set.
\end{enumerate}
The detection emerging from this process becomes the primary detection in the final catalogue.   The second best option is also retained and made available in the final catalogue with magnitudes labelled with a '2', i.e.: $i2$, $H\alpha2$, $r_I2$, $r_U2$, etc. as the secondary detection. A subset of the flags describing primary detections are provided for secondary detections also.  Not every primary detection is accompanied by a secondary detection.

Once two separate catalogues are created, one for IPHAS and one for UVEX,  with the selected primary and secondary detections in each, the two catalogues are merged, again using the {\sc tmatch} tool within {\sc stilts}.  Because stars vary, the cross-matching of the two catalogues does not insist on a maximum difference in $r$ magnitude before accepting -- accordingly, acceptance of a cross-match is based entirely on the astrometry.

\subsection{Compiling the final source list and advice on selection}
\label{sec:user-advice}

The final catalogue contains 174 columns, as described in the Appendix \ref{sec:AppTables}. In order to try to minimise spurious sources, we enforce two further cuts on the final catalogue:
\begin{enumerate}
    \item Objects with measurements in only the $U_{RGO}$ band are not included.
    \item A source should have a detection limit of S/N>5 in at least one of the other bands: i.e. it is required that at least one of $iErr$,  $haErr$, $rErr_I$,  $rErr_U$ or $gErr$ is smaller than 0.2\,mag.
\end{enumerate}

This leads to a final catalogue of $295.4\times10^6$ rows, each associated with a unique sky position.  This splits into $264.3/245.8\times10^6$ rows in which IPHAS/UVEX measurements are provided. Both IPHAS and UVEX photometric data are available for a subset of $214.7\times10^6$ objects.  Table~\ref{tab:stats} provides details on the numbers of sources for different combination of filters across the two surveys, together and separately. The number of stars raising no flags, for which errBits$=0$, are also given for each of the tabulated combinations.

In general terms, sources with detections in several bands are most likely real. However, there can also be real objects that are picked up in only one band. For example, very red and faint sources may have only a detection in $i$. Or a knot within a region of $H\alpha$ extended nebulosity, may appear in the catalogue as an $H\alpha$-only measurement.   Broadly speaking, we recommend reliance on the various warning flags available, and on the number of measurements $nObs_I$ and $nObs_U$ listed, in concluding on whether a source is real or spurious.   When the user wants to limit a selection to purely the best-quality detections over all available bands, the appropriate action is to include the requirement, $errbits=0$.

\begin{table}\centering
\begin{tabular}{lcc}
        & N ($\times10^6$)  &N ($\times10^6$) \\
        &                   &errBits=0\\
\hline
\multicolumn{3}{c}{IGAPS (surveys combined)}\\
\hline
All     &   295.4   &	205.2\\
IPHAS   &	264.3   & 	186.1\\
UVEX        &	245.8   &	170.7\\
IPHAS $+$ UVEX &   214.7   &  	151.6\\
\hline
\multicolumn{3}{c}{IPHAS}\\
\hline
$i,H\alpha,r_I$ & 168.4         & 115.4  \\
$i,r_I$&    31.7      & 25.2  \\
$i$           &   25.6       &  18.9 \\
$H\alpha$&    15.7      &  11.2 \\
$r_I$           &   16.3       &  12.0 \\
\hline
\multicolumn{3}{c}{UVEX}\\
\hline
$r_U,g,U_{RGO}$   &   54.3     &30.0   \\
$r_U,g$         &       101.1   &  72.7 \\
$r_U$           &       76.2   &  60.6 \\
$g$           &       12.7   &  6.8 \\
\hline
\end{tabular}
\caption{Number of sources in the catalogue for the stated survey/filter combinations. The first column of numbers counts all catalogue rows, while the second gives totals for the best quality errbits=0 sources.  Combinations of filters not shown individually account for less than 2\% of the total number of catalogue rows. The IPHAS part of the table pays no attention to whether there are any UVEX detections and vice versa for the UVEX part of the table.}
\label{tab:stats}
\end{table}

\section{Evaluation of the catalogue contents}
\label{sec:appraise}

\subsection{On photometric error as a function of magnitude} 

\begin{figure}\centering
 \resizebox{.9\hsize}{!}{\includegraphics{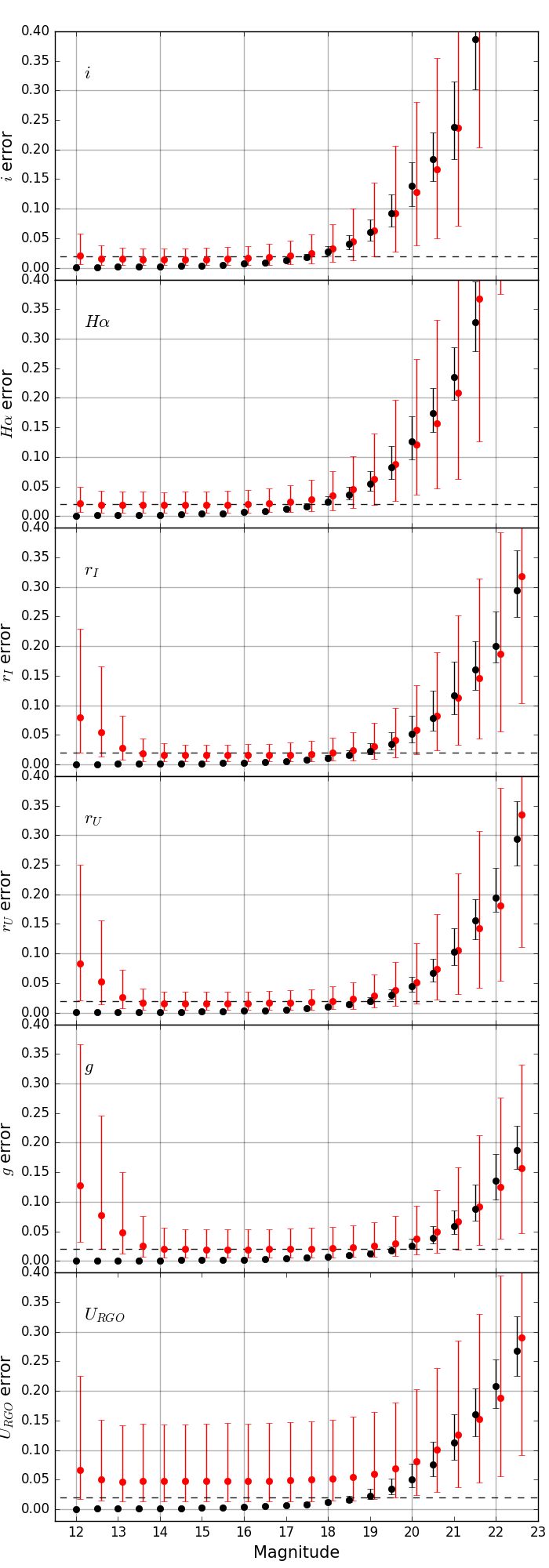}}
\caption{In black, median photometric errors for 0.5\,mag bins for each of the six bands. Error bars indicate the 16 and 84 percentiles, mimicking 1$\sigma$ error bars. In red we show the median differences between the primary and the secondary detections for each bin, with error bars indicating also their 16 and 84 percentiles. Red points are shifted 0.1\,mags to separate them from the black dots to make them visible. The dashed horizontal line marks a 0.02\,mag error level.}
\label{fig:meanerrors}
\end{figure}

\begin{figure*}\centering
 \resizebox{.9\hsize}{!}{\includegraphics{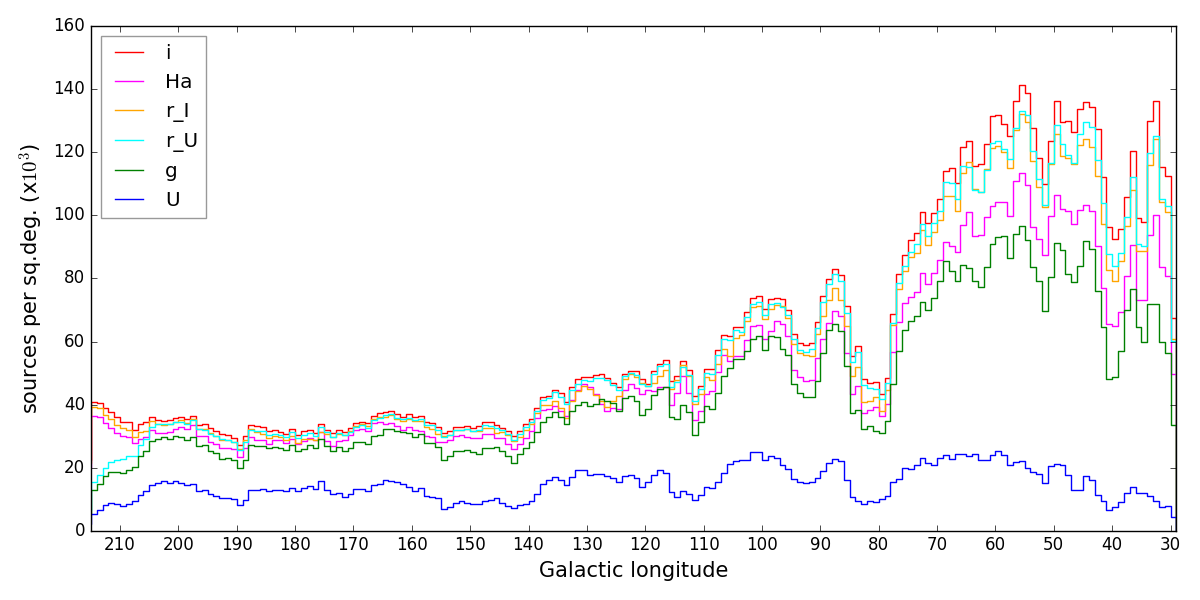}}
\caption{Number of sources as a function of Galactic longitude in each of the six pass bands, subject to the following requirements: $i < 20.5$ mag; the $i$ PSF is star-like (iClass < 0); errbits < 2 (see section~\ref{sec:user-advice}).}
\label{fig:numsources}
\end{figure*}

The median photometric errors reported in the catalogue are shown as a function of magnitude in figure~\ref{fig:meanerrors}.
They are assigned by the pipeline on the basis of the expected Poissonian noise in the aperture photometry.
In order to estimate the scale of the errors associated with their reproducibility (in effect, a scatter about the mean Pan-STARRS reference), we also plot in figure~\ref{fig:meanerrors} the absolute median magnitude difference between each primary detection and its corresponding secondary. We note that the secondary detection will, by definition, be lower quality in some aspect than the primary, and that the total number of measures available is smaller than the total number of primary detections because not every primary has a secondary. The error bars on both measures indicate the 16 and 84 percentiles of the errors for all the sources in a given 0.5\,mag bin. 

The effect of saturation is clear to see at the bright end in figure~\ref{fig:meanerrors} for the most sensitive $r$ and $g$ filters. Essentially the photometry worsens noticeably relative to results at fainter magnitudes at $r < 12.5$ and $g < 13.0$.  The very best photometry is achieved between 13.0 and 18-19 mag, depending on filter. In this range, reproducibility rather than random error dominates.   In all filters except $U_{RGO}$, the median error level is at or below 0.02, and shows more scatter than implied by the pipepline random error. This level has been drawn into figure~\ref{fig:meanerrors} to aid the eye. In $r$ it is between 0.015 and 0.02.  Factors contributing to the reproducibility error would include a mix of real data effects (e.g. focus gradients within the CCD footprint), and imprecisions in the data processing  (e.g. the dispersion around the adjustment of the illumination correction, known to be $\sigma$=0.008 -- see section \ref{sec:ilu}).

At faint magnitudes ($>$20th), the median primary-secondary differences are comparable with and can sometimes be lower than the Poissonian error.  The greater dispersion of the errors in $U_{RGO}$ band reflects at least in part the fact this band is not yet uniformly calibrated.

\subsection{On the numbers of sources by band and Galactic longitude}

Previous works based on the IPHAS survey alone have already investigated how the density of source detections in the $r$, $i$ and $H\alpha$ bands depends on Galactic longitude \citep{IPHASIDR,IPHASDR2,Farnhill2016}.  Of particular note in this regard is the study by \cite{Farnhill2016} which also looks at completeness in the $r$ and $i$ bands.  Here, we bring the added UVEX filters into view.

Figure~\ref{fig:numsources} shows the latitude-averaged density of all catalogued objects as a function of Galactic longitude for each of the six survey bands, subject to the requirement that a good detection in the $i$ band is available at a magnitude less than 20.5 (the median 5$\sigma$ limit - see figure~\ref{fig:5sig}). The effect of extinction is clear to see in that, in the first Galactic quadrant, even the $r$ stellar densities are a little lower than in $i$. The limiting magnitudes of the $H\alpha$ and $i$ data are much the same, and so the $H\alpha$ detection density is noticeably lower when extinction is more significant.  At all longitudes, the density of $U_{RGO}$ detected objects is between $\sim10$ and $\sim$20 thousand per sq.deg. ($\sim4$ per sq.arcmin.).  It is worth noting that, where $i < 18$, the detection rate in $g$, $r$ and $H\alpha$ relative to $i$ band is close to 100\%, and $\sim$50\% or better in $U_{RGO}$: as $i$ increases above 18, there is a progressive peeling away until the position shown in figure~\ref{fig:numsources} is reached.  In the second Galactic quadrant, there is good and quite even coverage in all bands (with $U_{RGO}$ at $\sim40$\%, all the way down to $i \sim 20$ mag.

The decrease in source density for the UVEX bands at Galactic longitude $\sim$210$\degr$ reflects the missing UVEX coverage in the corner of the footprint (see section \ref{sec:Obs}).

\subsection{On internal astrometric accuracy}\label{sec:difastro}
As described in section \ref{sec:merging} the cross match between bands was done in two steps, with a 1\,arcsec radius. In table \ref{tab:astrodif} we provide data on how this works out in practice: we compare the differences in astrometry between bands, based on the  $mDeltaRA$ and $mDeltaDEC$ catalogue columns for each band. We provide the median and 99 percentile separations for stars up to $r$<20 and also without any magnitude cut. 

The contemporaneous bands in IPHAS show typical astrometric differences that are consistent with the quality of the re-fit to the Gaia DR2 frame presented in section~\ref{sec:astro}.  The same is true for the contemporaneous UVEX $r_U$ vs $g$ separations. The cross match between the IPHAS and UVEX fields using the non-contemporaneous astrometry for the $r_I$ and $r_U$ bands gives slightly larger median values, but it is still the case that separations as large as 0.5 arcsec are extremely rare. The greatest difference is encountered when the $U_{RGO}$ filter is involved.  The median $r_U$ to $U_{RGO}$ separation of 0.1 arcsec is nevertheless broadly compatible with the residuals of the astrometry refit (cf. the bottom row of table~\ref{tab:astrodif} with the right panel of figure~\ref{fig:astdiff}).

\begin{table}\centering
\begin{tabular}{lcccc}
\hline
&\multicolumn{2}{c}{$r<20$}&\multicolumn{2}{c}{All sources}\\
percentiles             & 50\% & 99\% & 50\% & 99\%\\\hline
$r_I$ vs $i$            & 0.04  & 0.36  & 0.06  & 0.43  \\
$r_I$ vs $H\alpha$      & 0.04  & 0.38  & 0.06  & 0.45  \\
$r_I$ vs $r_U$          & 0.05  & 0.34  & 0.07  & 0.47  \\
$r_U$ vs $g$            &  0.04 & 0.36  & 0.06  & 0.43  \\
$r_U$ vs $U_{RGO}$      & 0.10  & 0.48  & 0.10  & 0.48  \\
\hline
\end{tabular}
\caption{Median and 99 percentile for the source position differences between bands. Units in arcseconds.}
\label{tab:astrodif}
\end{table}

\subsection{Comparison with Gaia and Pan-STARRS}

In order to compare our catalogue depth and completeness we have developed a simple unfiltered cross match with the Gaia DR2 catalogue \citep{2018A&A...616A...1G} in two 1 sq.deg. regions. The first is a high stellar density region: 60\degr$< \ell <$ 61\degr, 0\degr< b <1\degr, and the second one at  100\degr$< \ell <101$, -1\degr< b <0\degr is a lower density region. The cross match uses a wide 1\,arcsec radius, and keeps only the best option for each source. 

\begin{figure}\centering \resizebox{.8\hsize}{!}{\includegraphics{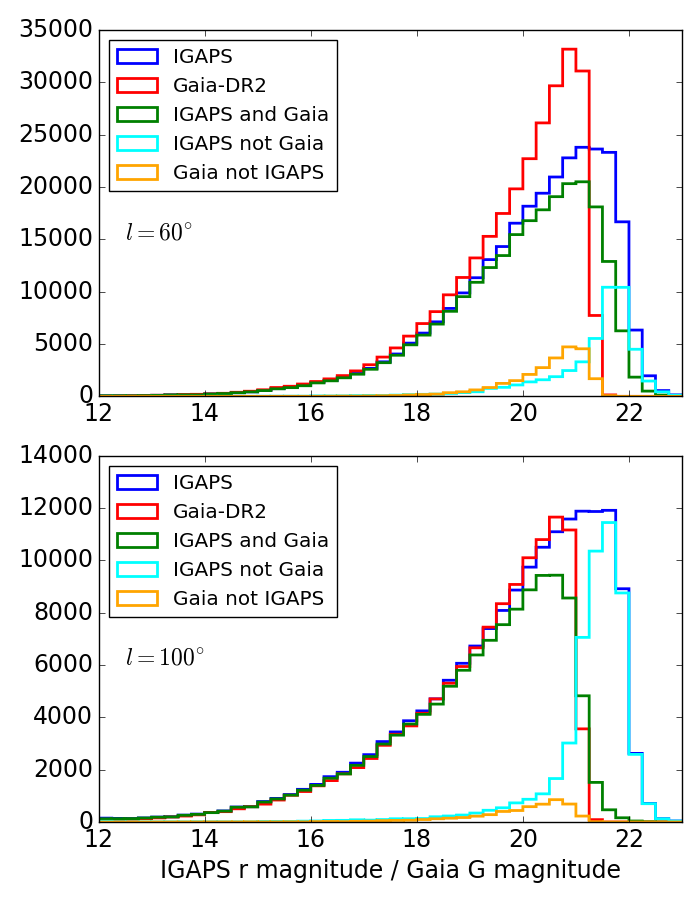}}
\caption{Results of the cross match between IGAPS and Gaia-DR2. Top: $\ell=60^{\circ}$, bottom: $\ell=100^{\circ}$. 
When, as is commonly the case, two $r$ magnitudes ($r_I$ and $r_U$) are available for an IGAPS source, the mean value is plotted.}
\label{fig:gaia}
\end{figure}

In both regions the total number of sources in IGAPS is larger than in Gaia. The reason for this can be seen in figure~\ref{fig:gaia}, where it is evident that the stars without Gaia counterparts are concentrated at fainter magnitudes, beyond Gaia's brighter limiting magnitude of $G=$20.5.  The small number of sources in Gaia but not in IGAPS (gold histogram) are spread in magnitude between $\sim$18th mag and the faint limit.  
 There are more of them at $\ell = 60^{\circ}$ than at $\ell = 100^{\circ}$, where there is undoubtedly more crowding. If the Gaia sources left unpaired by the initial match are cross-matched a second time with the IGAPS catalogue, then 9693/25286 at $\ell=60^{\circ}$ and 2156/5250 at $\ell=100^{\circ}$ find partners (already partnered in the first round) -- a $\sim$40\% success rate.  This behaviour shows that the much sharper Gaia PSF resolves more sources at faint magnitudes.  
At $\ell=60^{\circ}$ we have a density in the region of 300\,000 sources/sq.deg.  At a  a typical IGAPS seeing of 1-1.2\,$\arcsec$ (see figure~\ref{fig:5sig}), this leads to a $\sim$1/11 source per beam, well above the rule-of-thumb 1/30 confusion limit mentioned by \cite{Hogg01}.  At $\ell = 100^{\circ}$ the source density is lower by a factor of 2, roughly.

\begin{figure}\centering \resizebox{.9\hsize}{!}{\includegraphics{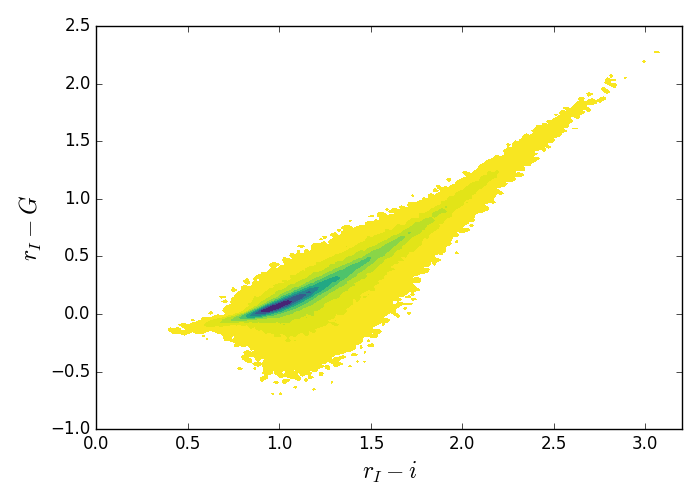}}
\caption{Differences between IGAPS $r_I$ and Gaia $G$ magnitudes as a function of IGAPS $r_I-i$ colour. The colour scales according to the density of sources in each bin, with square root intervals. Yellow represents the lowest density of at least 4 sources per 0.02x0.02\,mag$^2$ bin.}
\label{fig:rG}
\end{figure}

We have checked the quality flags for the sources found in IGAPS but not in Gaia to reject the hypothesis that they are just noise. 80\% of the sources not found in Gaia have $ErrBits$=0 making it unlikely they are spurious sources. Note that in figure~\ref{fig:gaia} we have as x axis both IGAPS $r$ and Gaia G magnitudes, which despite being very similar for modest $r-i$, have a growing colour dependence for increasing $r-i$, as can be seen in figure~\ref{fig:rG}.  A minor factor in figure~\ref{fig:gaia} is that IGAPS sources might not have a measured $r$ magnitude (either $r_I$ or $r_U$), and so could not be included.

\begin{figure}\centering
 \resizebox{.8\hsize}{!}{\includegraphics{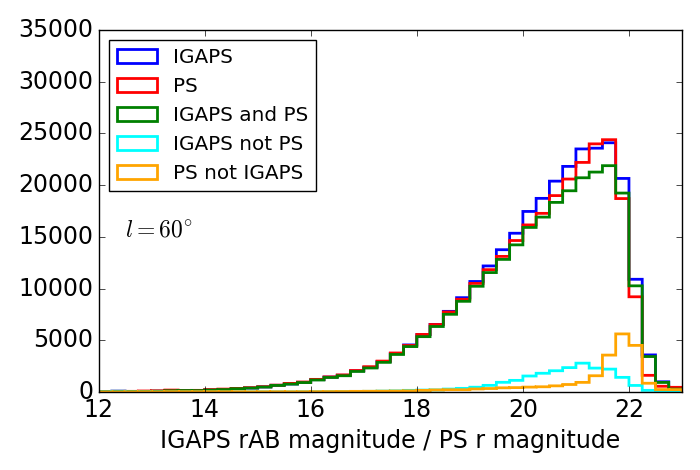}}
\caption{Results of the crossmatch between IGAPS and Pan-STARRS at $\ell=60\degr$. 
In blue, $rAB$ magnitude distribution for the IGAPS sources. In red, $r$ magnitude distribution from Pan-STARRS. In green, sources with both IGAPS and Pan-STARRS values. In cyan, sources in IGAPS not crossmatched with Pan-STARRS. In orange, sources in Pan-STARRS but not in IGAPS.
When two $r$ magnitudes are available for an IGAPS source ($rAB_I$ and $rAB_U$), then mean value is plotted, in a way that when one of them is missing, the source is plotted as well.}
\label{fig:ps}
\end{figure}

In the same two 1 sq.deg. areas, we have compared the IGAPS catalogue with Pan-STARRS \citep{Chambers16}.
In this case there are more sources in the Pan-STARRS catalogue. In figure~\ref{fig:ps} we can see that Pan-STARRS is a bit deeper in the $r$ band, but not by much. In this figure we are directly comparing Pan-STARRS and IGAPS AB magnitudes that are the same by construction.
Crowding accounts for less of the difference in this comparison since both catalogues come from ground-based photometric surveys with a similar pixel scale (0.333 vs 0.258 ''/pixel) and typical seeing.

\subsection{The fully calibrated colour-colour diagrams}

\begin{figure}\centering
 \resizebox{1.0\hsize}{!}{\includegraphics{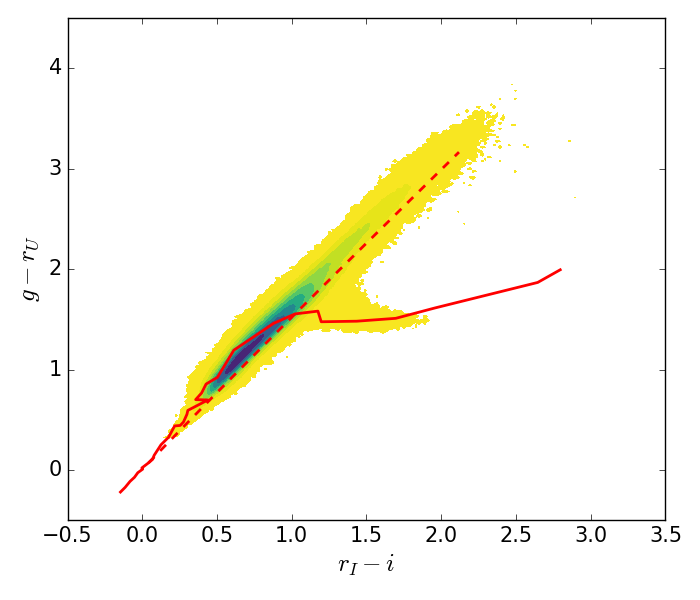}}
\caption{$g-r_U$ vs $r_I-i$ diagram for the longitude range, $60^{\circ}<\ell<65^{\circ}$. As in figure~\ref{fig:cc} the density of sources is portrayed by the squared root contoured colours, with yellow representing the lowest density of 4 sources per 0.02x0.02\,mag$^2$ bin. Note that the peak density traced by the darkest colour is over 5000 per bin.  Only sources with $r_I<19$ and errBits=0 have been used. The solid line in red is the unreddened main sequence, while the dashed line is the reddening line  for an A2V star up to $A_V$=10.}
\label{fig:grvri}
\end{figure}

\begin{figure}\centering
 \resizebox{.9\hsize}{!}{\includegraphics{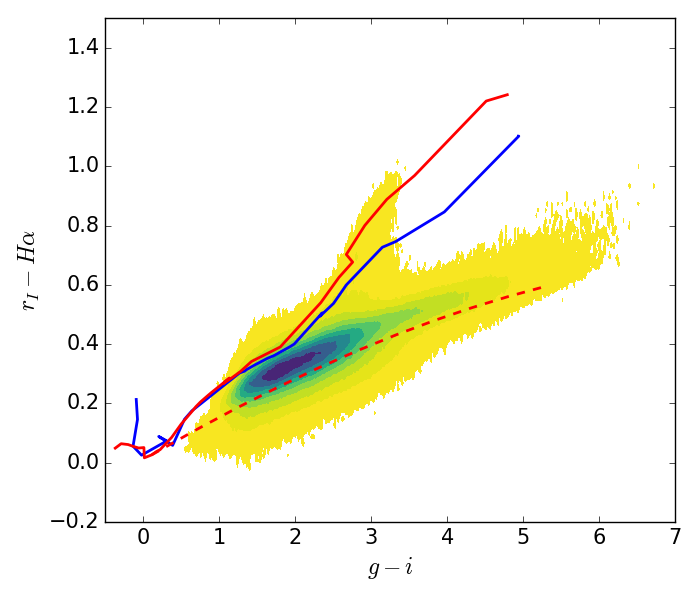}}
 \resizebox{.9\hsize}{!}{\includegraphics{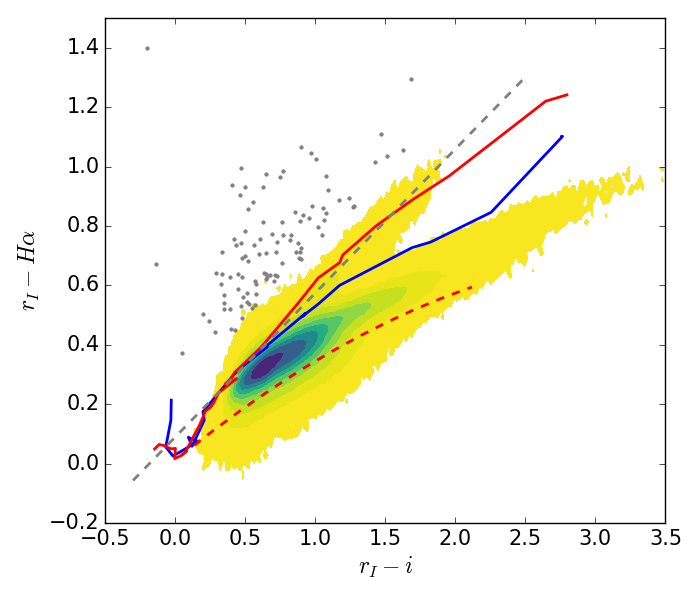}}
\caption{The two colour-colour diagrams involving $H\alpha$: in the top panel, $r_I-H\alpha$ vs $g-i$ and in the bottom, $r_I-H\alpha$ vs $r_I-i$, both shown for the Galactic longitude range, $60^{\circ}<\ell<65^{\circ}$. As in figure~\ref{fig:grvri}, the density of catalogued sources is portrayed by the squared root contoured colours.  Only sources with $r_I<19$ and errBits=0 have been plotted. The solid line in red is the unreddened main sequence, while the dashed line is the  reddening line for an A2V star up to $A_V$=10. The blue line is the sequence for the giants.  The grey dashed line is the emitters selection cut appropriate to these longitudes, applied within the range $-0.3<r_I-i<2.5$, while the grey dots are the selected emitters at $>5\sigma$. The emitter selection is presented in section~\ref{sec:emitter}.}
\label{fig:ccHa}
\end{figure}

The creation of the IGAPS catalogue adds to the available colour-colour diagrams.  The first of these to mention is the $g-r_U$, $r_I-i$ diagram that uses the three fully calibrated broad bands. An example, constructed as a density plot from the Galactic longitude range $60^{\circ} < \ell < 65^{\circ}$, is shown in figure~\ref{fig:grvri}.  The tracks overplotted in red have been computed via synthetic photometry using library spectra (see Appendix D).
As the main sequence (MS) and giant tracks sit very nearly on top of each other, we show only the MS track as a red solid line.
A reddening line  for an A2V star is also included as a dashed line. The comparison of the catalogue data with these reference tracks points out that all stars to K-type fall within a neat linear strip that follows the reddening vector.  Only the M stars break away from this trend, creating the roughly horizontal thinly-populated spur at $g - r_U \sim 1.5$ where nearly unreddened M dwarfs are located.  This can be echoed at greater $g-r_U$ and $r_I - i$ by an even sparser distribution of stars to the right of the main stellar locus. Indeed, in the example shown in Figure~\ref{fig:grvri} it happens the density of stars is too low to be visible. Stars in this region will be mainly reddened M giants. Similarly, a thin scatter of points below the unreddened M-dwarf spur and redwards of the main locus can occur. These will be white dwarf -- red dwarf binaries \citep{Augusteijn08}.

There are two fully-calibrated colour-colour diagrams now available that involve $r_I - H\alpha$, the available measure of $H\alpha$ excess.  Our examples of them, in figure~\ref{fig:ccHa}, come from the same longitude range as shown for $g-r_U$ vs $r_I - i$ (figure~\ref{fig:grvri}).  Using $g - i$ as the abscissa (top panel in the figure) naturally offers a much greater numeric range than is possible when $r_I - i$ is used instead (bottom panel).  The important difference in form between them is that in the $g - i$ diagram, the unreddened MS track turns through an angle in the M-star domain creating a spur above the main run of the stellar locus, in which increasing interstellar extinction drags the main stellar locus to the right and up only $\sim0.2$ in $r_I-H\alpha$ over $\Delta(g-i) \sim 3$.  The unreddened giant track (shown in blue) does not change angle quite as much as the MS track and yet remains quite close to it.  As a result, the part of the diagram redward of the unreddened M-type spur and above the main locus will be occupied by a mix of reddened red giants and some candidate (reddened) emission line stars.   

In the $r_I - i$ diagram the dwarf and giant M stars smoothly continue the trend line established in the FGK range, and there is more separation between them.  This means a little less of the colour-colour space falls between the M-type main sequence and the domain dominated by  giant stars.  This means more of the stars located in this gap are likely to be emission line stars than in the case of the diagram using $g-i$.  Practically, these differences confer some advantage on the $r - H\alpha$ vs $r_I - i$ diagram for the selection of emission line stars.  

\subsection{The $U_{RGO}$ filter data and the UVEX colour-colour diagram}
\label{sec:URGO}

A problem in the calibration of all $U$-like filters with transmission extending into the ground-based ultraviolet is that the effective band pass is weather-dependent.  Worse still, \cite{Patat2011} have shown that weather shifts in the atmosphere influencing ultraviolet throughput are uncorrelated with changes at wavelengths greater than 400\,nm.  The combination of this behaviour with the fact of high and variable extinction in Galactic Plane fields represents a calibration challenge that is best met by an astrophysical method, such as that described by \cite{MS-II}.  We have not attempted this here.  So far, there is in place a pipeline adjustment that imposes a fixed offset between the $U_{RGO}$ and $g$ filter zero points, which amounts to a preliminary relative calibration.

\begin{figure}\centering
 \resizebox{.9\hsize}{!}{\includegraphics{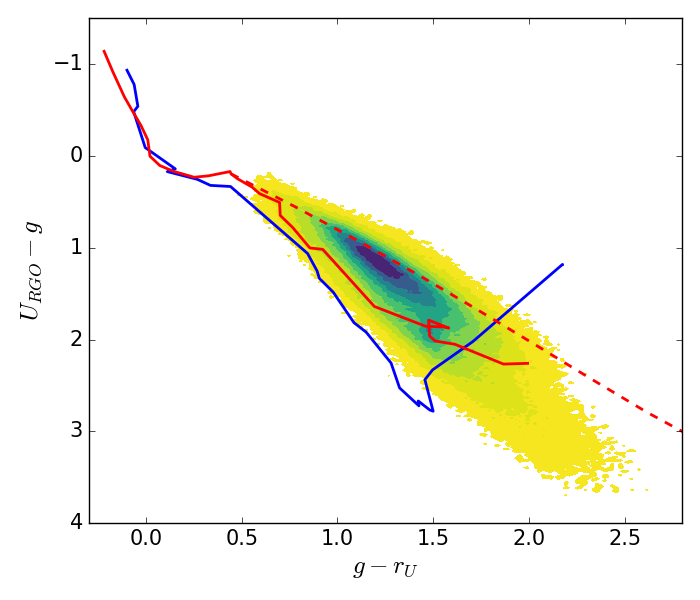}}
 \resizebox{.9\hsize}{!}{\includegraphics{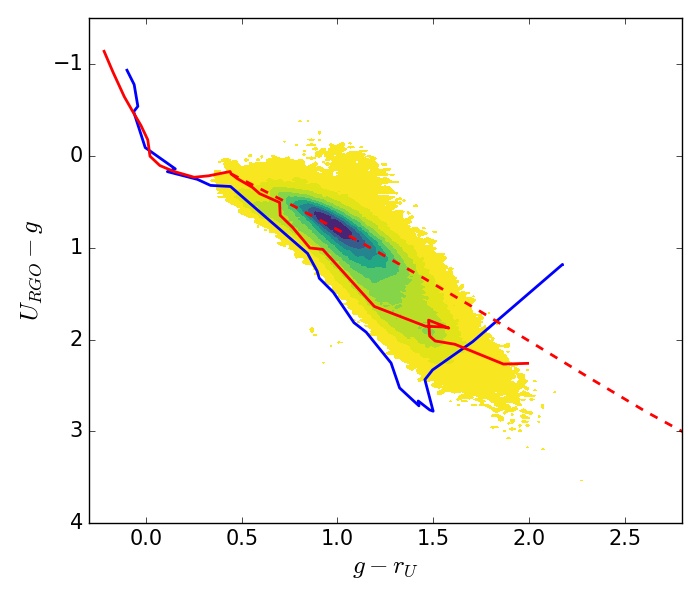}}
\caption{$U_{RGO}-g$ vs $g-r$ diagram for the regions $80^{\circ}<l<85^{\circ}$ (top) and $185^{\circ}<l<190^{\circ}$ (bottom). The density of sources is portrayed by the contoured colour scale.   Sources with $g<20$ and errBits=0 are included. The solid line in red in both is the unreddened main sequence, while the dashed line is the reddening line  for and F5V star extended up to $A_V$=10.  The blue line represents the giants. Numerical detail on the tracks is provided in Appendix D.  The top panel is an example of a region in which the pipeline calibration has produced a uniform outcome, while the lower panel provides an instance of where it is clear that there is some variation in the $U_{RGO}$ photometric scale.}
\label{fig:ugvgr}
\end{figure}

Two examples of how the preliminary calibration works out is shown in figure~\ref{fig:ugvgr}.  Since both $g$ and $r_U$ are globally calibrated, only photometric offsets in $U_{RGO}$ will disturb the main stellar locus.  The upper panel of figure~\ref{fig:ugvgr} provides an instance of a region within the catalogue (in Cygnus) where there is evidence of a stable $U_{RGO}$ photometric scale: the main stellar locus has the expected properties and indeed is quite well aligned with the run of the F5V reddening line and the lower bound set by the giant track \citep[for more on the expected behaviour and the impact of red leak, see][]{Verbeek2012}.  In contrast the lower panel is an example of a part of the outer Galactic disc, observable during the winter months from La Palma, when spells of photometric stability are less common.  This is signalled by the outlier islands of data points above and below the main stellar locus.  Even here, it is evident that much of the region shares a consistent $U_{RGO}$ calibration (if a little bright, judging by the reddening line that slices through the region of peak stellar density, when it should sit on top of it).  

An obvious astrophysical difference between the two colour-colour diagrams in figure~\ref{fig:ugvgr} is the greater extension of the red, i.e. lower-right, tail in Cygnus as compared with the outer disc.  This betrays the greater extinction and the presence of more red giants to be expected at the lower Galactic longitude. A striking feature of the unreddened giant track is the nearly right-angles turn as the latest M types are reached.  It has the consequence that, redwards of $g - r_U \sim 2$, M8--10 giants will co-locate with O and early B stars, where the latter are reddened by more than $\sim$ 8 visual magnitudes. 

As things stand, the $U_{RGO}$ magnitudes included in the catalogue can be regarded as subject to a relative calibration that may not be too far from an absolute one.  Hence, the value of the magnitudes provided is that they are well-suited to first-cut discrimination of UV-bright or UV-excess sources with respect to the stellar fields in which they are embedded.   

\section{Applications of the data release}\label{sec:applications}

We focus on just two applications that enable two further columns in the released catalogue, each picking out group of objects of specific astrophysical interest.  These groups are candidate emission line stars and variable stars with $r$ magnitude differences greater than 0.2 mag.

\subsection{The selection of emission line stars from the IPHAS $(r_I - H\alpha)$ versus $(r_I  - i)$ diagram}\label{sec:emit}
\label{sec:emitter}

The IPHAS survey on its own supports one colour-colour diagram and this has been discussed extensively in previous works \citep{IPHAS, Sale09, IPHASDR2}. The two important utilities of $(r_I-H\alpha)$ versus $(r_I-i)$ are the means to separate spectral type from extinction for many stars \citep{Sale09}, and to identify candidate emission line objects \citep[hereafter W08]{Witham08}.  

The method of identification of emission line stars is to pick out objects with $(r_I - H\alpha)$ colour greater than that of unreddened main sequence (MS) stars of the same $(r_I - i)$ -- given that the impact of non-zero extinction on MS stars is to displace their positions in the diagram rightward and upward along a trajectory running below the unreddened sequence.  Selection above the MS locus can produce a highly reliable, if incomplete, list of candidate emission line stars.

The first effort to do this was presented by W08 on the basis of what was then an incomplete and not-yet-calibrated IPHAS database.  The outcome was a list of 4853 candidate emission line stars down to a limiting magnitude of $r = 19.5$, dependent on a selection process working with $r$, $i$ and $H\alpha$ data at the level of individual fields.  Follow up spectroscopy in the Perseus Arm has since indicated a low rate of contamination at magnitudes down to $r\sim17$ \citep{Raddi15, Gkouvelis16}.  We revisit the selection, taking advantage of the survey-wide uniform calibration of $r$, $i$ and $H\alpha$ now available.  

Like W08 we only search for emission line stars at $r < 19.5$ mag.  We remove from consideration any star for which its image in any band is classified as 'noise-like' (morphology class 0). We also reject any star for which any warning flag is raised in any IPHAS band, with the exception that we  permit a bright neighbour.   We do not require the existence of a second detection confirming an $H\alpha$ excess.  In this last respect the selection is unlikely to reject emission line objects that also vary rapidly.  The defining step of the selection is to measure the $(r-H\alpha)$ colour excess relative to a reference line of fixed slope that emulates the trend of the mean observed main sequence.  The reference line takes the form,
\begin{equation}
\label{eqn:hacut}
r_I-H\alpha = 0.485(r_I-i) + k(\ell)
\end{equation}
and it is only applied over the range $-0.3 < r_I-i < 2.5$.  

In equation~\ref{eqn:hacut}, $k(\ell)$ is a constant varying slowly with Galactic longitude, that is intended to track the height of the mean main sequence above the $r_i - H\alpha = 0$ axis.  We have noticed a small but definite modulation with Galactic longitude such that $k(\ell)$ peaks at $\sim 0.09$ at $\ell \simeq 80^{\circ}$ and in the third quadrant, declines to a minimum of $\sim 0.06$ near $\ell = 150^{\circ}$. The likely cause is the longitude dependence of the amount of extinction located within a few hundred parsecs of the Sun \citep[see Figures 9 to 11 in][]{Lallement2019}: essentially, when extinction builds up quickly over the first few hundred pc the main sequence locus in the ($r-H\alpha,r-i$) diagram shifts a little toward increased $r-i$ (lowering $k(\ell)$).  We capture this with a piecewise fit made up of three linear segments tracking this variation:  
\begin{equation}
    \begin{aligned}
    k(\ell) & = 0.0706 + 2.8754\times10^{-4}\ell \ \ & (\ell < 77^{\circ}.90) \\
            & = 0.1303 - 4.7874\times10^{-4}\ell \ \ & (77^{\circ}.90 < \ell < 150^{\circ}.22) \\
            & = -0.0378 + 6.4031\times10^{-4}\ell \ \ & (\ell > 150^{\circ}.22)
    \end{aligned}
\end{equation}
The rms scatter of the offsets about this function -- determined from 74 $5\times5$ sq.deg. samples spanning the catalogue -- is 0.0076.   An example of the cut line and its longitude-sensitive placement can be seen in figure \ref{fig:ccHa} showing the Galactic longitude range $60 ^{\circ}< \ell < 65^{\circ}$.

\begin{figure}\centering
\resizebox{.9\hsize}{!}{\includegraphics{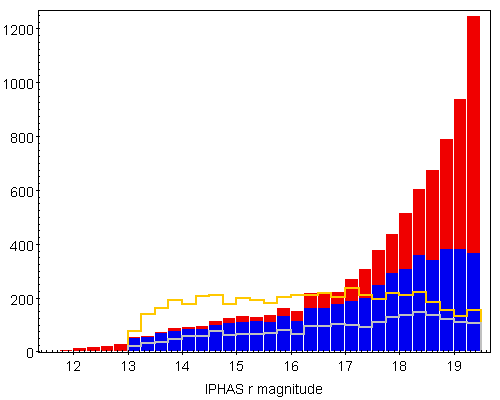}}
\caption{A comparison between the r magnitude distributions of $>5\sigma$ candidate emission lines stars identified in the IGAPS catalogue and the W08 list.  The red filled histogram refers to the full IGAPS list, while the superimposed blue filled histogram is limited to objects meeting the same morphology-class criteria imposed by W08. The yellow unfilled histogram represents the full W08 list. The light grey unfilled histogram shows the union of the full W08 list with the IGAPS list (blue histogram), when restricted to candidate emitters meeting W08's class criteria and bright limit.
}\label{fig:emitter_r}
\end{figure}

\begin{figure*}\centering
 \resizebox{.9\hsize}{!}{\includegraphics{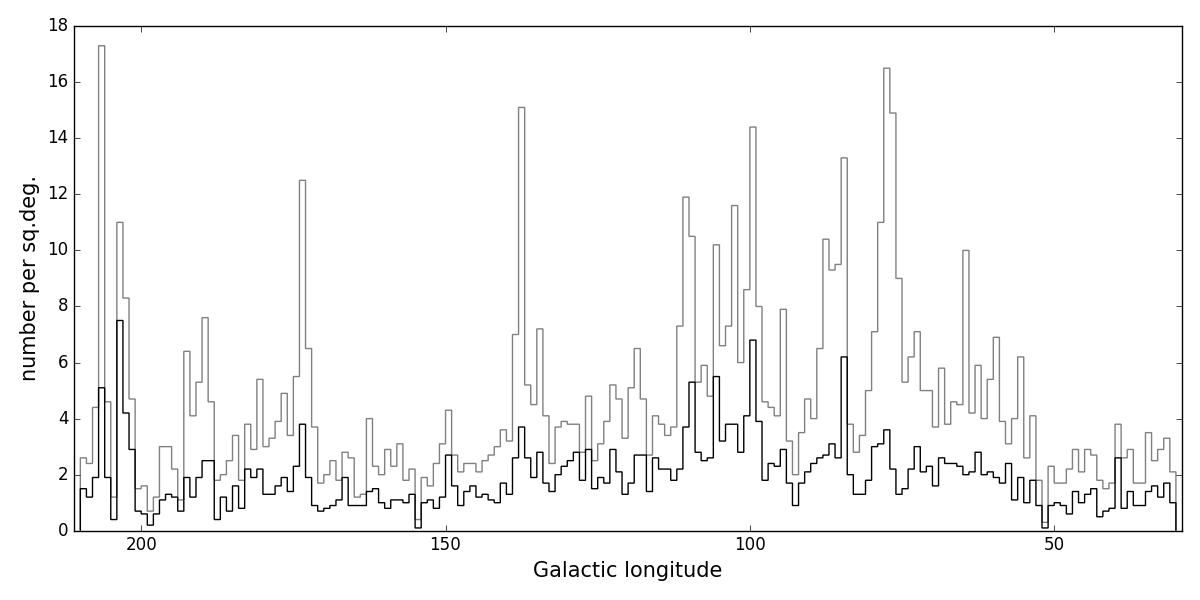}}
\caption{Distribution of candidate emission line objects as a function of Galactic longitude. The grey histogram incorporates all objects marked as 2 in the $emitter$ catalogue column. These are stars with $H\alpha$ excess greater than 5$\sigma$ and $r_I$<19.5. The black histogram is limited to those sources with $r_I$<18.}\label{fig:emittersl}
\end{figure*}

\begin{figure}\centering
\resizebox{.95\hsize}{!}{\includegraphics{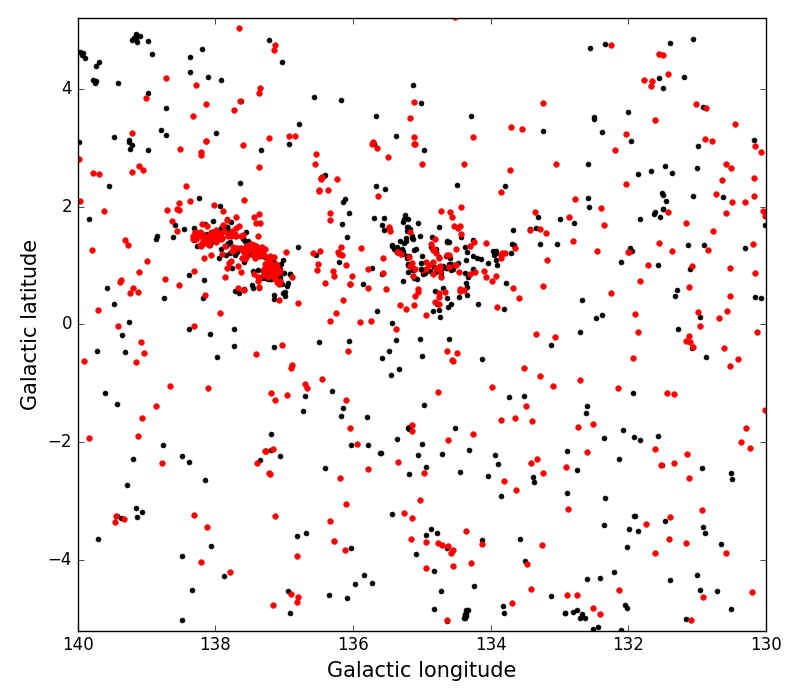}}
\caption{Distribution of the candidate emitters in the $10\times10$ sq.deg. box containing the Heart and Soul nebulae at respectively $\ell \simeq 135^{\circ}$ and $\ell \simeq 138^{\circ}$, in the Perseus Arm. High-confidence emitters selected with excess greater than 5$\sigma$ are in red. More marginal candidates with excesses of between 3 and 5$\sigma$ are in black.
}\label{fig:emitters130}
\end{figure}

For a source to be accepted as a high-probability emission line star the vertical difference between its $r_I-H\alpha$ colour and the reference line needs to exceed 5$\sigma$, where the definition of $\sigma$ is:
\begin{equation}\label{eq:sig}
\sigma^2 = \sigma_{int}^2 + \epsilon_{H\alpha}^2+(1-m)^2\epsilon_{r_I}^2+m^2\epsilon_i^2
\end{equation}
where $m = 0.485$ is the gradient from equation~\ref{eqn:hacut}.
The first term in the quadrature sum is included in order to capture the intrinsic spread in $(r-H\alpha)$ at fixed $(r_I-i)$, plus an allowance for the reproducibility error in the photometry.  Each of these contributions is estimated to introduce scatter at a level of up to 0.02 (adding in quadrature to place a minimum total $\sigma$ of 0.028 at magnitudes brighter than $r \sim 16$).  The other terms are the appropriately-weighted individual-band random errors per source, as given in the IGAPS catalogue.   

A feature of this selection is that the required excess of $5\sigma$ will usually translate to a minimum $H\alpha$ emission equivalent width in the region of $\sim10$~\AA\ for bright stars ($r < 16$) with small random errors. 
This minimum can rise to over $\sim$30~\AA\ as $\sigma$ from equation \ref{eq:sig} trends towards $\sim 0.1$ for reddened objects at the faint end of the included magnitude range.  

The results of the new selection have been placed in an additional column named $emitter$ in the IGAPS catalogue. A number 2 is recorded where a source is found to be an emission line candidate at greater than the $5\sigma$ level, while the number 1 appears for marginal candidates in the 3$\sigma$--5$\sigma$ range.  A zero is recorded when the excess is $<3\sigma$ (or negative).  The entry is null if the test was not applied -- we chose not to apply it to very blue ($r-i < -0.3$) and very red ($r-i > 2.5$) stars because the cut applied has no meaning in these extreme domains. There are relatively few objects outside these limits.

Our $>5\sigma$ selection contains 8292 stars, while a further 12568 fall into the 3$\sigma$--5$\sigma$ group.  We have created a subset of the $>5\sigma$ candidates that satisfy the additional constraints imposed by W08.  These are that $r_I > 13$ and that the PSF is star-like (requires a morphology class < 0).  The 8292 stars are reduced to 4755 by this means, revealing that the excluded 3537 objects must be classified as extended (morphology class $+1$) in one or more IPHAS filters. Indeed, for a majority of the excluded stars, the narrowband  $H\alpha$ classification is $+1$.  It is apparent in figure~\ref{fig:emitter_r} that these are preferentially fainter than $r_I \sim 17.5$.  There is certainly a risk at fainter magnitudes that the sky subtraction of the $H\alpha$ flux is compromised in regions of pronounced and locally variable nebulosity, and may appear more extended as a result.

Another point to note is that, of the 4755 candidates meeting the additional W08 criteria, only $\sim45$\% are in common with the W08 list. 
Bringing into the statistics the cross-matched 3--5$\sigma$ stars makes little difference -- indeed a smaller fraction of them overlaps the W08 list.  How this has happened is suggested by figure~\ref{fig:emitter_r}: at magnitudes brighter than $r\sim16$ the new selection finds systematically fewer objects than W08 while, at $r > 17.5$ roughly, this turns round such that the new selection finds more.  Our treatment of the errors is likely to be more conservative at bright magnitudes than in W08's treatment (where the dominant term is the first in Equation~\ref{eq:sig}), and potentially less so at the faint end.     

Insight into the spatial distribution of candidate emitters is provided in figures~\ref{fig:emittersl} and \ref{fig:emitters130}.  The general features of the overall longitude distribution have much in common with a figure 3 of W08.  Once again, the sharp peaks line up with well known star-forming regions -- a point underlined by figure~\ref{fig:emitters130}, which shows how the Heart and Soul nebulae are well-populated with emission line stars.  In plotting the complete list in figure~\ref{fig:emittersl}, we have split it into two magnitude ranges such that the upper lighter grey histogram includes all $>5\sigma$ candidates down to $r_I = 19.5$, while the lower black histogram is limited to objects with $r_I < 18$.  The most nebulous part of the northern Galactic Plane is in the Cygnus-X region, running from around $\ell = 70^{\circ}$ to $\ell = 85^{\circ}$. It coincides with the domain in which there is, seemingly, a preponderance of fainter candidate emission line objects.  This is where we would expect there to be the most contamination of the emitters list at faint magnitudes due to uncertain sky subtraction in $H\alpha$.  

A full understanding of the properties and reliability of the new list of candidate emission line stars, has to come from confirmatory spectroscopy.  A useful feature of the approach we have taken is that it is fully-specified and thus entirely reproducible -- and it is easy to adapt.  A more exhaustive finer-grained approach, examining the position of the cut line in the colour-colour plane on a much smaller angular scale than the $5\times5$ sq.deg. used here, is recommended for the study of limited regions.

\subsection{Insights on stellar variability from the two r-magnitude epochs}

\begin{figure}\centering
 \resizebox{.99\hsize}{!}{\includegraphics{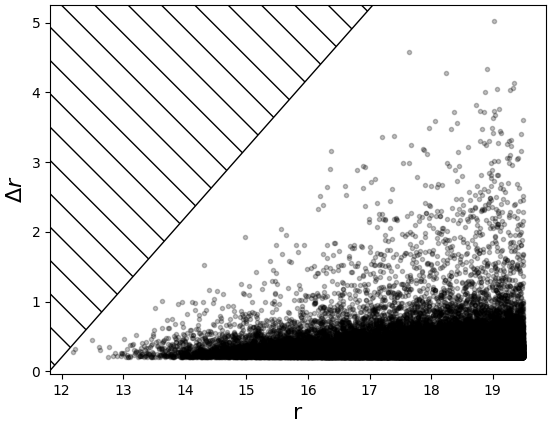}}
\caption{Distribution of candidate variable objects as a function of $r$ magnitude. The abscissa is the numerically greater of the two available $r$ magnitudes, $r_I$ and $r_U$.  The hatched area to top left is inaccessible to IGAPS given the bright limit of the merged catalogue.}
\label{fig:varrdr}
\end{figure}

\begin{figure}\centering
\resizebox{.99\hsize}{!}{\includegraphics{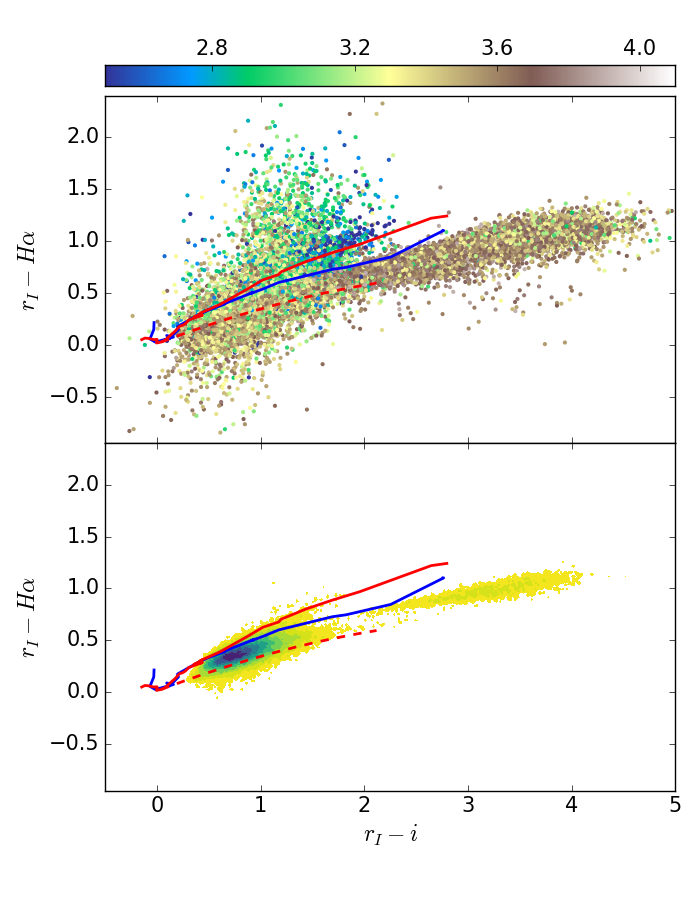}}
\caption{The $r_I-H\alpha$ vs $r_I-i$ two-colour diagram for variable sources. The stars in the upper panel are coloured according to the logarithm of the distance from GAIA DR2 \citep{2018AJ....156...58B}. The solid red line is the unreddened main sequence, while the dashed red line  is the reddening line for an A2V star up to $A_V$=10. The blue line is the unreddened sequence for the giants.  The lower panel is a density plot of the stars identified as variables.  Evidently the great majority are located in the main stellar locus at $r_I-i \lesssim 1$.  The other notable feature is the `island' of relatively extreme red giants beginning at $r_I-i \sim 2.5$.}
\label{fig:varriha}
\end{figure}

In the last decades many dedicated digital surveys for stellar variability, either from ground or space, have been conducted. Often these surveys avoid the Galactic Plane due to problems with crowding. Hence, while the IGAPS surveys were not designed to look for variability, they still might be used to detect variable sources if they happen to show a large enough variation between repeat observations. Repeat observations happen either due to a re-observation of a field due to bad data quality, in the overlaps between fields and offset fields, or due to the r-band filter purposely being used in both the IPHAS and UVEX surveys. Since observations repeated within either IPHAS or UVEX generally include one bad observation and field-pair overlaps are mostly observed just a few minutes apart, we concentrate here on the repeat observations in the $r$ band between the two surveys.

A star is flagged as variable in the catalogue if the absolute value of the difference between $r_I$ and $r_U$ exceeds 0.2 mag, and is larger than 5 times the combined photometric error of the two measurements plus a 0.015 mag systematic error (see figure~\ref{fig:meanerrors}).  We also require both $r$ magnitudes to be brighter than 19.5, that the source PSF is not noise-like in either measurements, and that the errBits cumulative flag is $<2$.

This selection leads to 53525 objects being classified as variable.  These are flagged in the $variable$ column in the catalogue. Figure~\ref{fig:varrdr} shows the distribution of change in magnitude versus the fainter magnitude of the object. Clearly a very large amplitude can only be found for objects that are detected towards the faint end of the range in one of the measurements, as they otherwise would be saturated or undetectable in the other measurement. The mean change in magnitude for the variables is 0.340, while the maximum is just over 5 mag. The mean time difference between observations is 1941.3 days, the minimum is 83 minutes and the maximum 5530.9 days. The majority of objects can be found at $r_I - i < 2$ (84\%) while 10\% of the objects are extremly red objects at $r_I - i > 3$.  Only 278 of the objects identified this way are listed in the General Catalogue of Variable Stars \footnote{http://www.sai.msu.su/gcvs/gcvs/} \citep{2017ARep...61...80S}. 125 of them are Miras, semiregular and irregular late type variables, while 63 objects are classified as eclipsing binaries, 35 as young variables, 18 as dwarf novae, 17 as pulsating variables.

There are 9 sources that show a magnitude change greater than 4. Three of these are listed as Mira or candidate Mira in SIMBAD. 8 out of them are very red with $r_I - i \gtrsim 3$, and hence we would expect these to be Miras or semiregular variables. The final source turns out to be a nearby high proper motion star that happens to fall on top of a faint background star in one of the epochs, hence leading it to be wrongly classified as variable in unusual circumstances.

51292 sources have counterparts in the GAIA DR2 distance catalogue \citep{2018AJ....156...58B} within 0.5 arcsec. These are plotted in figure~\ref{fig:varriha} in the IPHAS two-colour diagram, where the data points are coloured according to the logarithm of the distance in parsecs. Nearer-by stars are predominantly at $r_I - i < 2$ and show $H\alpha$ in emission. A lot of these sources are likely to be young stellar objects (YSOs), while the closest of all (coloured deep blue in the figure) will be active M dwarfs. The furthest stars, at distances of a few kiloparsecs, are mostly found at $2 < r_I - i < 3$ (the darkest brown points in the upper panel of Fig.\,\ref{fig:varriha}). It is likely these are giants in sightlines with relatively little interstellar extinction. Stars that are redder than $r_I - i > 3$ seem to be a bit closer, suggesting that these extreme red colours are associated with more circumstellar or interstellar reddening. 

Sources falling below the red-dashed  reddening line for an A2V star in the plot often present problems in their measurements due to either unflagged bad pixels or large background variations created by bright neighbours or nebulosity. Hence, many in this modest group of $\sim$ 500 sources are likely to be interlopers. But not all: for example, some of the reddest in this domain may be genuine carbon stars \citep[see Section 6.3 in][]{IPHAS}. The lower panel of figure~\ref{fig:varriha} confirms that they all sit in a part of the colour-colour plane that is very thinly populated.  This is also true of the stars lying above the unreddened main sequence.  Indeed, the number of stars in common between the `variable' and `emitter' categories is 1219.  YSOs will dominate this group.  Finally, we note that
21 variables have $r - H\alpha > 2$. 7 of them are classified in SIMBAD as YSOs, 3 as symbiotics and 2 as PN.

\section{Closing remarks}\label{Concl}

The main goal of this paper has been to present the new IGAPS catalogue, formed from merging the IPHAS \citep{IPHAS} and UVEX \citep{UVEX} surveys of the northern Galactic Plane.  It is a catalogue of 174 columns and almost 300 million rows, spanning the $r$ magnitude range from 12--13th mag down to 21st (10$\sigma$, see figure~\ref{fig:meanerrors}). The astrometry in all 5 photometric bands has been placed in the Gaia DR2 reference frame. Broadband $g$, $r$, and $i$ have been uniformly calibrated using Pan-STARRS data resting on that project's 'Ubercal' \citep{Magnier2013}. We estimate the reproducibility of the photometry in these bands (and in $H\alpha$) to be in the region of 0.02 magnitudes at magnitudes brighter than $\sim$19th.  

The key diagnostic bands in IGAPS are narrow-band $H\alpha$ (IPHAS) and the $u$ band as mimicked by the $U_{RGO}$ filter (UVEX).  The large number of sources available per exposure in $H\alpha$ has made possible a uniform calibration across the full 1850 sq.deg. footprint. In a follow up publication presenting the database of IGAPS images (Greimel et al, in prep.) we will use this to set a flux scale to the $H\alpha$ images, so they may be fully exploited in studies of extended nebulae and the ionized interstellar medium.  Here, we directly use the $H\alpha$ calibration in identifying a list of candidate emission line stars: these number 8292 at $>5\sigma$ significance down to a faint limit $r_I = 19.5$.  The challenge of the much lower source density in the $U_{RGO}$ exposures has meant that the calibration so far remains as computed on a run-by-run basis in the pipeline processing.  This has turned out to be reasonably stable, if more approximate.  It is adequate e.g. for the selection of stars with UV excess.  

The UVEX and IPHAS surveys both obtained data in the $r$ band, at two distinct epochs that are typically several years apart.  Both epochs are given in the IGAPS catalogue and have been used to make a global selection of stellar variables brighter than $r = 19.5$, subject to a threshold, $|\Delta r| > 0.2$. The total found implies roughly 1 in 4000 catalogued objects are, by this definition, significant variables.  

This first federation of UVEX blue photometry with red IPHAS data provides, in the IGAPS catalogue, a resource of great utility for the examination of the northern Milky Way's stellar content.  Previous applications of the separate survey databases have ranged all the way from local white dwarfs up to the most luminous and massive stars detected at heliocentric distances of up to 10 kpc. The merger, especially now that increasingly precise astrometry is flowing from the Gaia mission as well, can become a convenient basis for more flexible and incisive analysis of early, late and high-mass stellar evolution.  An immediate use will be in the selection of Galactic Plane targets for the upcoming WEAVE spectroscopic survey on the William Herschel Telescope \citep{WEAVE}. 
The IGAPS catalogue will be made world-accessible via the Centre de Donn\'ees Astronomique (CDS) in Strasbourg.

\begin{acknowledgements}
This work is based on observations made with the Isaac Newton Telescope operated on the island of La Palma by the Isaac Newton Group of Telescopes in the Spanish Observatorio del Roque de los Muchachos of the Instituto de Astrof\'isica de Canarias. We would like to take this opportunity to thank directly Marc Balcells (ING Director), Cecilia Farina, Neil O'Mahony, Javier M\'endez, and other members of ING staff who have lent their support to this programme of work over the years, helping to bring it to the finishing line.\\
MM, JED and GB acknowledge the support of research grants  funded  by  the  Science,  Technology  and  Facilities Council of the UK (STFC, grants ST/M001008/1 and ST/J001333/1).
MM was partially supported by the MINECO (Spanish Ministry of Economy) through grant ESP2016-80079-C2-1-R and RTI2018-095076-B-C21 (MINECO/FEDER, UE), and MDM-2014-0369 of ICCUB (Unidad de Excelencia 'Mar\'ia de Maeztu'). 
RG benefitted from support via STFC grant ST/M001334/1 as a visitor to UCL.
PJG acknowledges support from the Netherlands Organisation for Scientific Research (NWO), in contributing to the Isaac Newton Group of Telescopes and through grant 614.000.601. 
JC acknowldges support by the Spanish Ministry of Economy, Industry and Competitiveness (MINECO) under grant AYA2017-83216-P.
DJ acknowledges support from the State Research Agency (AEI) of the Spanish Ministry of Science, Innovation and Universities (MCIU) and the European Regional Development Fund (FEDER) under grant AYA2017-83383-P.
RR acknowledges funding by the German Science foundation (DFG) through grants HE1356/71-1 and IR190/1-1.\\
We thank Eugene Magnier for providing support on Pan-STARRS data.
This research has made use of the University of Hertfordshire high-performance  computing  facility  (https://uhhpc.herts.ac.uk/) located at the University of Hertfordshire (supported by STFC grants including ST/P000096/1). We thank Martin Hardcastle for his support and expertise in connection with our use of the facility.\\
This work has made use of data from the European Space Agency (ESA) mission
{\it Gaia} (\url{https://www.cosmos.esa.int/gaia}), processed by the {\it Gaia}
Data Processing and Analysis Consortium (DPAC,
\url{https://www.cosmos.esa.int/web/gaia/dpac/consortium}). Funding for the DPAC
has been provided by national institutions, in particular the institutions
participating in the {\it Gaia} Multilateral Agreement.
Much of the analysis presented has been carried out via {\sc TopCat} and {\sc stilts} \citep{2006ASPC..351..666T}.\\
We thank the referee for comments on this paper that have improved its content.
\end{acknowledgements}

%
%

   \bibliographystyle{aa} 
   \bibliography{IGAPSAA} 

\begin{appendix} 
\onecolumn
\section{Details on the exposure grading system}
\label{tab:grades}

This table expands on the information about the quality checks on the individual-field exposure sets discussed in section~\ref{sec:quality}. 
\bigskip

\begin{center}
\begin{tabular}{ll}
\hline
Grade & Requirements \\
\hline
A++ & Seeing$\leq$ 1.25$\arcsec$ \\
   & standard deviation wrt Pan-STARRS $stdps$<0.04 \\
   &  \\
A+ & Seeing$\leq$ 1.5$\arcsec$ \\
   & standard deviation wrt Pan-STARRS $stdps$<0.04 \\
   & \\
A  & Seeing$\leq$ 2.0$\arcsec$ \\
    & standard deviation wrt Pan-STARRS $stdps$<0.04 \\
    & \\
B   & Seeing$\leq$ 2.5$\arcsec$ \\
    & standard deviation wrt Pan-STARRS $stdps<0.05$ \\
 & \\
C  &Seeing$\leq$ 2.5$\arcsec$ \\
   & standard deviation wrt Pan-STARRS $stdps$<0.08 \\
   & \\
D  &  Seeing $>$ 2.5$\arcsec$ \\
   &  or ellipticity $>$ 0.3\\ 
   & or number of stars for Pan-STARRS comparison < 100\\
   & or limiting magnitude ($5\sigma$): $i$>19, $H\alpha$>19, $r$>20, $g$>20\\
  & or moon separation <20$\degr$\\
  &  or strong photometric difference in $H\alpha$ within field pair \\
  &\,\,\,\,\,\,\,($> 98$ percentile for total field distribution).\\
  & or manually graded as D through visual inspection of the image.\\
\hline  
  \end{tabular}
  \end{center}

\newpage

\section{Placement of the $g$ filter mask}
\label{sec:gmask_supp}

\begin{figure*}[h!]\centering
 \resizebox{.85\hsize}{!}{\includegraphics{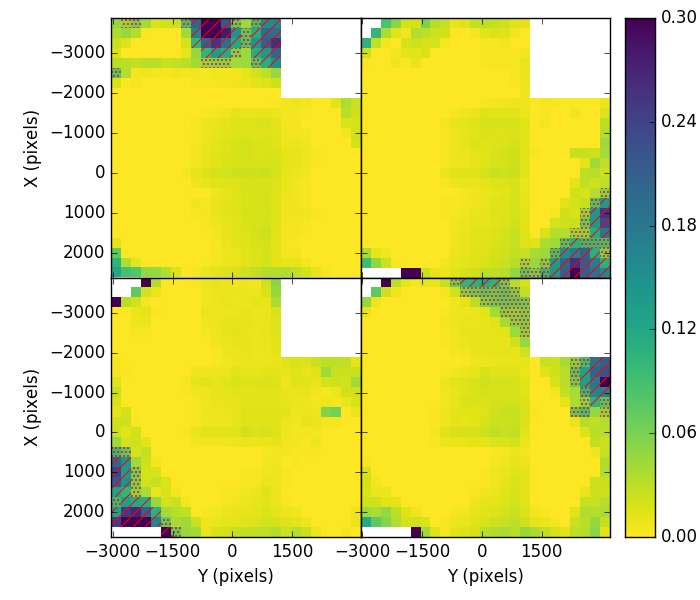}}
\caption{Differences between Pan-STARRS and IGAPS $g$ magnitudes as a function of position in the WFC image plane. Median values are plotted for each 250x250 pixel$^2$ bin. The mask applied for observations made within four different phases of UVEX data collection  are shown.  The diagonal hatched regions represent the placement of the inner $g$-band mask, while the dotted regions indicate the outer mask. Top-left: mask used for observations before June 2006.
Top-right: mask for observations between June 2006 and December 2013. Bottom-left: mask for observations between December 2013 and March 2017. Bottom-right: mask for observations after March 2017.}
\label{fig:gmask}
\end{figure*}

In section~\ref{sec:gband} the impact of a blemish on the $g$ band filter used in the execution of UVEX was described, along with its mitigation.  We show how the mask for flagging affected $g$ magnitudes was applied to the data in figure~\ref{fig:gmask}.  When the $g$ filter was cleaned and replaced in its mount, it did not always go back in oriented as before.  Indeed in the late stages of observation, an effort was made to try to re-orient the filter so that the blemish would fall in front of the cut-out corner of the detector array.  

\newpage
\onecolumn

\section{Catalogue Columns}\label{sec:AppTables}
\begin{longtable}[h!]{llcl}
\caption{List of columns available in the catalogue, together with the units and brief description of the column content.
}
\\\label{tab:Table1}
No & Column & Units & Description \\ \hline
1	&	name	&		&Source designation (JHHMMSS.ss+DDMMSS.s) without IGAPS prefix.	\\
2	&	RA	&deg		&J2000 RA (Gaia DR2 reference frame).	
\\
3	&	DEC	&	deg	&J2000 DEC (Gaia DR2 reference frame).
\\
4	&	gal\_long	&deg	&	Galactic Longitude.	\\
5	&	gal\_lat	&deg	&   Galactic Latitude.	\\
6	&	sourceID	&	&	Unique source identification string (run-ccd-detectionnumber).\\
7	&	posErr	&	arcsec	&Astrometric fit error (rms) across the CCD.\\
8	&	mergedClass	&	&	1=galaxy, 0=noise, -1=star, 99=if different filters don't agree. See sect.\,\ref{sec:flags}.\\
9	&	pStar	&		&Probability that the source is stellar.	\\
10	&	pGalaxy	&		&Probability that the source is extended.\\
11	&	pNoise	&		&Probability that the source is noise.	\\
12	&	i	&	mag	&IPHAS i mag (Vega) using the 2.3 arcsec aperture.\\
13	&	iErr	&		mag&Random uncertainty for i. When r is not available and no colour term has been  \\
 & & & used, 0.05 mag has been added in quadrature.	\\
14	&	iAB	&	mag	&IPHAS i mag (AB) using the 2.3 arcsec aperture.	\\
15	&	iEll	&		&Ellipticity in the i-band.	\\
16	&	iClass	&		&1=galaxy, 0=noise, -1=star, -2=probableStar, -3=probableGalaxy for the i band.\\
17	&	iDeblend	&	&	True if the i source is blended with a nearby neighbour.	\\
18	&	iSaturated	&	&	True if the i source is saturated.	\\
19	&	iVignetted	&	&	True if the i source is in a part of focal plane where there is vignetting.	\\
20	&	iTrail	&		&True if the i source is close to a linear artifact.\\
21	&	iTruncated	&	&	True if the i source is close to the CCD boundary.\\
22	&	iBadPix	&		&True if there are bad pixel(s) in the i source aperture.\\
23	&	iMJD	&		&Modified Julian Date at the start of the i-band exposure.	\\
24	&	iSeeing	& arcsec		&Average FWHM in the i-band exposure.	\\
25	&	iDetectionID	&	&Unique i-band detection identifier (run-ccd-detectionnumber).\\
26	&	iDeltaRA	&	arcsec	&Position offset of the i-band detection in RA.\\
27	&	iDeltaDEC	&	arcsec	&Position offset of the i-band detection in DEC.\\
28	&	ha	&	mag	&IPHAS H-alpha mag (Vega) using the 2.3 arcsec aperture.	\\
29	&	haErr	&mag		&Random uncertainty for ha. \\	
30	&	haAB	&	mag	&IPHAS ha mag (AB) using the 2.3 arcsec aperture.	\\
31	&	haEll	&		&Ellipticity in ha band.\\
32	&	haClass	&		&1=galaxy, 0=noise, -1=star, -2=probableStar, -3=probableGalaxy for the ha band.\\
33	&	haDeblend	&	&	True if the ha source is blended with a nearby neighbour.	\\
34	&	haSaturated	&	&	True if the ha source is saturated.	\\
35	&	haVignetted	&	&	True if the ha source is in a part of focal plane where  there is vignetting.	\\
36	&	haTrail	&		&True if the ha source is close to a linear artifact.\\
37	&	haTruncated	&	&	True if the ha source is close to the CCD boundary.\\
38	&	haBadPix	&	&	True if there are bad pixel(s) in the ha source aperture.\\
39	&	haMJD	&		&Modified Julian Date at the start of the ha exposure.	\\
40	&	haSeeing	&arcsec	&	Average FWHM in the ha exposure.	\\
41	&	haDetectionID	&	&	Unique ha detection identifier (run-ccd-detectionnumber).\\
42	&	haDeltaRA	&	arcsec	&Position offset of the ha-band detection in RA.	\\
43	&	haDeltaDEC	&	arcsec	&Position offset of the ha-band detection in DEC.\\
44	&	r\_I	&	mag	&IPHAS r mag (Vega) using the 2.3 arcsec aperture.\\
45	&	rErr\_I	&	mag	&Random uncertainty for r\_I. \\	
46	&	rAB\_I	&	mag	&IPHAS r mag (AB) using the 2.3 arcsec aperture.\\
47	&	rEll\_I	&		&Ellipticity in r\_I.\\
48	&	rClass\_I	&	&	1=galaxy, 0=noise, -1=star, -2=probableStar, -3=probableGalaxy for the r\_I band.		\\
49	&	rDeblend\_I	&	&	True if the r\_I source is blended with a nearby neighbour.		\\
50	&	rSaturated\_I	&	&	True if   the r\_I source is saturated.	\\
51	&	rVignetted\_I	&	&	True if  the r\_I source is in a part of focal plane where  there is vignetting.		\\
52	&	rTrail\_I	&	&	True if the r\_I source is close to a linear artifact.	\\
53	&	rTruncated\_I	&	&	True if the r\_I source is close to the CCD boundary.\\
54	&	rBadPix\_I	&	&	True if there are bad pixel(s) in the r\_I source aperture.\\
55	&	rMJD\_I	&		&Modified Julian Date at the start of the r\_I exposure.\\
56	&	rSeeing\_I	&arcsec	&	Average FWHM in the r\_I exposure.\\
57	&	rDetectionID\_I	&	&	Unique r\_I detection identifier (run-ccd-detectionnumber).\\
58	&	r\_U	&	mag	&UVEX r mag (Vega) using the 2.3 arcsec aperture.	\\
59	&	rErr\_U	&	mag	&Random uncertainty for r\_U. \\	
60	&	rAB\_U	&	mag	&UVEX r mag (AB) using the 2.3 arcsec aperture.		\\	
61	&	rEll\_U	&		&Ellipticity in r\_U.		\\
62	&	rClass\_U	&	&	1=galaxy, 0=noise, -1=star, -2=probableStar, -3=probableGalaxy for the r\_U band.		\\
63	&	rDeblend\_U	&	&	True if the r\_U source is blended with a nearby neighbour.		\\	
64	&	rSaturated\_U	&	&	True if the r\_U source is saturated.		\\	
65	&	rVignetted\_U	&	&	True if the r\_U source is in a part of focal plane where  there is vignetting.		\\	
66	&	rTrail\_U	&	&	True if the r\_U source is close to a linear artifact.		\\	
67	&	rTruncated\_U	&	&	True if the r\_U is close to the CCD boundary.		\\	
68	&	rBadPix\_U	&	&	True if there are bad pixel(s) in the r\_U source aperture.		\\	
69	&	rMJD\_U	&		&Modified Julian Date at the start of the r\_U exposure.		\\
70	&	rSeeing\_U	&arcsec	&	Average FWHM in the r\_U exposure.		\\	
71	&	rDetectionID\_U	&	&	Unique r\_U detection identifier (run-ccd-detectionnumber).		\\	
72	&	rDeltaRA\_U	&	arcsec	&Position offset of the r\_U detection in RA.		\\	
73	&	rDeltaDEC\_U	&	arcsec	&Position offset of the r\_U detection in DEC.		\\	
74	&	g	&	mag	&UVEX g mag (Vega) using the 2.3 arcsec aperture.		\\	
75	&	gErr	&	mag	&Random uncertainty for g. When r is not available  and no colour term has been  \\
 & & & used, 0.05 mag has been added in quadrature.	\\	
76	&	gAB	&	mag	&UVEX g mag (AB) using the 2.3 arcsec aperture.		\\	
77	&	gEll	&		&Ellipticity in the g-band.		\\	
78	&	gClass	&		&1=galaxy, 0=noise, -1=star, -2=probableStar, -3=probableGalaxy for the g band.		\\
79	&	gDeblend	&	&	True if the g source is blended with a nearby neighbour.		\\	
80	&	gSaturated	&	&	True if the g source is saturated.		\\	
81	&	gVignetted	&	&	True if the g source is in a part of focal plane where  there is vignetting.		\\	
82	&	gTrail	&		&True if the g source is close to a linear artifact.		\\	
83	&	gTruncated	&	&	True if the g source is close to the CCD boundary.		\\	
84	&	gBadPix	&		&True if there are bad pixel(s) in the g source aperture.		\\	
85	&	gmask	&		&Source located in the inner (1) or outer (2) degraded area in the g-band filter.		\\
86	&	gMJD	&		&Modified Julian Date at the start of the g-band exposure.		\\	
87	&	gSeeing	&arcsec		&Average FWHM in the g-band exposure.		\\	
88	&	gDetectionID	&	&	Unique g-band detection identifier (run-ccd-detectionnumber).		\\	
89	&	gDeltaRA	&	arcsec	&Position offset of the g-band detection in RA.		\\	
90	&	gDeltaDEC	&	arcsec	&Position offset of the g-band detection in DEC.		\\	
91	&	U\_RGO	&	mag	&UVEX U\_RGO mag (Vega) using the 2.3 arcsec aperture.  Default pipeline calibration.		\\	
92	&	UErr	&	mag	&Random uncertainty for U\_RGO.  Pipeline random error only.		\\	
93	&	UEll	&	mag	&Ellipticity in U\_RGO band.		\\	
94	&	UClass	&		&1=galaxy, 0=noise, -1=star, -2=probableStar, -3=probableGalaxy for the  U\_RGO band.		\\
95	&	UDeblend	&	&	True if the U\_RGO source is blended with a nearby neighbour.		\\	
96	&	USaturated	&	&	True if the U\_RGO source is saturated.		\\	
97	&	UVignetted	&	&	True if the U\_RGO source is in a part of focal plane where  there is vignetting.		\\	
98	&	UTrail	&		&True if the U\_RGO is close to a linear artifact.		\\	
99	&	UTruncated	&	&	True if the U\_RGO is close to the CCD boundary.		\\	
100	&	UBadPix	&		&True if there are bad pixel(s) in the U\_RGO source aperture.		\\	
101	&	UMJD	&		&Modified Julian Date at the start of the U\_RGO exposure.		\\	
102	&	USeeing	&	arcsec	&Average FWHM in the U\_RGO exposure.		\\	
103	&	UDetectionID	&	&	Unique U\_RGO detection identifier (run-ccd-detectionnumber).		\\	
104	&	UDeltaRA	&	arcsec	&Position offset of the U\_RGO-band detection in RA.		\\	
105	&	UDeltaDEC	&	arcsec	&Position offset of the U\_RGO-band detection in DEC.	\\	
106	&	brightNeighb	&		&True if a very bright star is nearby.		\\	
107	&	deblend	&		&True if the source is blended with a nearby neighbour in one or more bands.		\\	
108	&	saturated	&		&True if saturated in one or more bands.		\\	
109	&	nBands	&		&Number of bands in which the source is detected.		\\
110	&	errBits	&		&Bitmask indicating: bright neighbour (1), source blending (2), trail (4), saturation (8), \\
&&&outer gmask (16), vignetting (64), inner gmask (128), truncation (256) \\
&&&and bad pixels (32768).		\\
111	&	nObs\_I	&		&Number of repeat IPHAS observations of this source.		\\
112	&	nObs\_U	&		&Number of repeat UVEX observations of this source.		\\
113	&	fieldID\_I	&	&	Survey field identifier in IPHAS, e.g. 0001, 0001o, 0002, etc.		\\	
114	&	fieldID\_U	&	&	Survey field identifier in UVEX, e.g. 0001, 0001o, 0002, etc.		\\	
115	&	fieldGrade\_I	&	&	Internal quality control score of the IPHAS field. A to D scale.		\\	
116	&	fieldGrade\_U	&	&	Internal quality control score of the UVEX field. A to D scale. 		\\	
117 & emitter & & 2 if good candidate for $H\alpha$ line emission, 1 if marginal, 0 if tested and in main locus,\\
&&& null if not tested.\\
118 & variable & & True if difference between the IPHAS and UVEX $r$ measurements exceeds \\
&&&5$\sigma$ and 0.2 mag.\\
119	&	2SourceID	&	&	SourceID of the object in the second detection.		\\	
120	&	i2	&	mag	&IPHAS i mag (Vega) for the secondary detection.		\\	
121	&	i2Err	&	mag	&Random uncertainty for i2. When r2 is not available  and no colour term has been used,  \\
 & & &0.05 mag has been added in quadrature.	\\	
122	&	i2Class	&		&1=galaxy, 0=noise, -1=star, -2=probableStar, -3=probableGalaxy for the i2 band.		\\	
123	&	i2Seeing	&arcsec	&	Average FWHM in the i2 exposure.		\\	
124	&	i2MJD	&		&Modified Julian Date at the start of the i2 exposure.		\\	
125	&	i2DeltaRA	&	arcsec	&Position offset of the i2-band detection in RA.	\\	
126	&	i2DeltaDEC	&	arcsec	&Position offset of the i2-band detection in DEC.		\\	
127	&	i2DetectionID	&		&Unique i2 detection identifier (run-ccd-detectionnumber).		\\	
128	&	i2ErrBits	&		&Bitmask indicating bright neighbour (1), source blending (2), trail (4), saturation (8),  \\
&&&vignetting (64),  truncation (256) and bad pixels (32768) for i2.		\\
129	&	ha2	&	mag	&IPHAS H-alpha mag (Vega) for secondary detection.		\\	
130	&	ha2Err	&	mag	&Random uncertainty for  ha2.		\\	
131	&	ha2Class	&		&1=galaxy, 0=noise, -1=star, -2=probableStar, -3=probableGalaxy for the ha2 band.		\\	
132	&	ha2Seeing	&	arcsec	&Average FWHM in the ha2 exposure.		\\	
133	&	ha2MJD	&		&Modified Julian Date at the start of the ha2 exposure.		\\	
134	&	ha2DeltaRA	&	arcsec	&Position offset of the ha2-band detection in RA.		\\	
135	&	ha2DeltaDEC	&	arcsec	&Position offset of the ha2-band detection in DEC.		\\	
136	&	ha2DetectionID	&		&Unique ha2 detection identifier (run-ccd-detectionnumber).		\\	
137	&	ha2ErrBits	&		&Bitmask indicating bright neighbour (1), source blending (2), trail (4), saturation (8),  \\
&&&vignetting (64),  truncation (256) and bad pixels (32768) for ha2.		\\
138	&	r2\_I	&	mag	&IPHAS r mag (Vega) for the secondary detection.		\\	
139	&	r2Err\_I	&	mag	&Random uncertainty for r2\_I.		\\	
140	&	r2Class\_I	&		&1=galaxy, 0=noise, -1=star, -2=probableStar, -3=probableGalaxy for the r2\_I band.		\\	141	&	r2Seeing\_I	&	arcsec	&Average FWHM in the r2\_I exposure.		\\	
142	&	r2MJD\_I	&		&Modified Julian Date at the start of the r2\_I exposure.		\\	
143	&	r2DeltaRA\_I	&	arcsec	&Position offset of the r2\_I-band detection in RA.		\\	
144	&	r2DeltaDEC\_I	&	arcsec	&Position offset of the r2\_I-band detection in DEC.		\\	
145	&	r2DetectionID\_I	&		&Unique r2\_I detection identifier (run-ccd-detectionnumber).		\\	
146	&	r2ErrBits\_I	&		&Bitmask indicating bright neighbour (1), source blending (2), trail (4), saturation (8),  \\
&&&vignetting (64),  truncation (256) and bad pixels (32768) for r2\_I.		\\
147	&	r2\_U	&	mag	&UVEX r mag (Vega) for the secondary detection.		\\	
148	&	r2Err\_U	&mag		&Random uncertainty for r2\_U.		\\	
149	&	r2Class\_U	&		&1=galaxy, 0=noise, -1=star, -2=probableStar, -3=probableGalaxy for the r2\_U band.		\\	
150	&	r2Seeing\_U	&	arcsec	&Average FWHM in the r2\_U exposure.		\\	
151	&	r2MJD\_U	&		&Modified Julian Date at the start of the r2\_U exposure.		\\	
152	&	r2DeltaRA\_U	&	arcsec	&Position offset of the r2\_U-band detection in RA.		\\	
153	&	r2DeltaDEC\_U	&	arcsec	&Position offset of the r2\_U-band detection in DEC.	\\	
154	&	r2DetectionID\_U	&		&Unique r2\_U detection identifier (run-ccd-detectionnumber).		\\	
155	&	r2ErrBits\_U	&		&Bitmask indicating bright neighbour (1), source blending (2), trail (4), saturation (8),  \\
&&&vignetting (64),  truncation (256) and bad pixels (32768) for r2\_U.		\\
156	&	g2	&	mag	&UVEX g mag (Vega) for the secondary detection.		\\	
157	&	g2Err	&mag		&Random uncertainty for  g2. When r2 is not available  and no colour term has been used,  \\
 & & &0.05 mag has been added in quadrature.	\\
158	&	g2Class	&		&1=galaxy, 0=noise, -1=star, -2=probableStar, -3=probableGalaxy for the g2 band.		\\	
159	&	g2Seeing	&arcsec	&	Average FWHM in the is exposure.		\\	
160	&	g2MJD	&		&Modified Julian Date at the start of the g2 exposure.		\\	
161	&	g2DeltaRA	&	arcsec	&Position offset of the g2-band detection in RA.		\\	
162	&	g2DeltaDEC	&	arcsec	&Position offset of the g2-band detection in DEC.		\\	
163	&	g2DetectionID	&		&Unique g2 detection identifier (run-ccd-detectionnumber).		\\	
164	&	g2ErrBits	&		&Bitmask indicating bright neighbour (1), source blending (2), trail (4), saturation (8),   \\
&&& outer gmask (16), vignetting (64), inner gmask (128),  truncation (256) and \\
&&&bad pixels (32768) for g2.		\\
165	&	U\_RGO2	&	mag	&UVEX U\_RGO mag (Vega) for the secondary detection.  Default pipeline calibration.			\\	
166	&	U2Err	&	mag	&Random uncertainty for  U\_RGO2.		\\	
167	&	U2Class	&		&1=galaxy, 0=noise, -1=star, -2=probableStar, -3=probableGalaxy for the U\_RGO2 band.		\\	
168	&	U2Seeing	&arcsec	&	Average FWHM in the U\_RGO2 exposure.		\\	
169	&	U2MJD	&		&Modified Julian Date at the start of the U\_RGO2 exposure.		\\	
170	&	U2DeltaRA	&	arcsec	&Position offset of the U\_RGO2-band detection in RA.		\\	
171	&	U2DeltaDEC	&	arcsec	&Position offset of the U\_RGO2-band detection in DEC.		\\	
172	&	U2DetectionID	&		&Unique U\_RGO2 detection identifier (run-ccd-detectionnumber).	\\	
173	&	U2ErrBits	&		&Bitmask indicating bright neighbour (1), source blending (2), trail (4), saturation (8),  \\
&&&vignetting (64),  truncation (256) and bad pixels (32768) for U\_RGO2.		\\
174	&	errBits2	&		&Global bitmask for the second detection indicating: bright neighbour (1), source blending (2), \\ 
&&&trail (4), saturation (8),  outer gmask (16), vignetting (64), inner gmask (128),\\
&&&  truncation (256) and bad pixels (32768).		\\
\end{longtable}

\newpage

\section{Tracks}\label{sec:AppTracks}

Synthetic colours for main sequence and giant stars, computed by folding spectra from the INGS spectral library, accessible at https://lco.global/$\sim$apickles/INGS/ , with the ING measured filter curves and an atmosphere calculated with ESO SkyCalc \citep{2012A&A...543A..92N, 2013A&A...560A..91J} for La Silla (similar altitude to La Palma), an airmass of 1.2 \citep[as used by Pan-STARRS,][and close to our survey median of 1.15]{2012ApJ...750...99T} and a precipitable water vapour (PWV) content of 5\,mm \citep{2009SPIE.7475E..1HG}. Optical surfaces are not taken into account, as precise measurements of them were not available.
The extinction law used is from \cite{1999PASP..111...63F}.
The full tables can be downloaded from the CDS.

\begin{table}[ht!]
\caption{Synthetic colour of selected dwarf stars for $R_V=3.1$}
\begin{tabular}{ccccccccccccc}
\hline
$Spectral$ & \multicolumn{4}{c}{$A_V = 0$} & \multicolumn{4}{c}{$A_V = 2$}  & \multicolumn{4}{c}{$A_V = 4$}  \\
$Type$ & $U_{RGO}-g$ & $g-r$ & $r-H\alpha$ & $r-i$ & $U_{RGO}-g$ & $g-r$ & $r-H\alpha$ & $r-i$ & $U_{RGO}-g$ & $g-r$ & $r-H\alpha$ & $r-i$ \\
\hline
\hline
B0 & -1.14 & -0.22 & 0.05 & -0.15 & -0.43 & 0.47 & 0.22 & 0.30 & 0.32 & 1.13 & 0.35 & 0.78 \\
B3 & -0.64 & -0.11 & 0.06 & -0.08 & 0.05 & 0.57 & 0.22 & 0.36 & 0.78 & 1.23 & 0.35 & 0.84 \\
B5 & -0.49 & -0.07 & 0.05 & -0.05 & 0.19 & 0.61 & 0.22 & 0.39 & 0.92 & 1.26 & 0.34 & 0.87 \\
B8 & -0.33 & -0.03 & 0.05 & -0.03 & 0.35 & 0.65 & 0.21 & 0.41 & 1.07 & 1.31 & 0.34 & 0.89 \\
A0 & 0.00 & 0.02 & 0.02 & -0.00 & 0.67 & 0.70 & 0.18 & 0.44 & 1.38 & 1.35 & 0.30 & 0.92 \\
A2 & 0.11 & 0.07 & 0.03 & 0.04 & 0.77 & 0.75 & 0.18 & 0.48 & 1.49 & 1.39 & 0.31 & 0.97 \\
A5 & 0.17 & 0.15 & 0.04 & 0.08 & 0.85 & 0.81 & 0.20 & 0.52 & 1.58 & 1.45 & 0.32 & 1.00 \\
F0 & 0.22 & 0.33 & 0.13 & 0.18 & 0.92 & 0.98 & 0.28 & 0.61 & 1.66 & 1.62 & 0.39 & 1.10 \\
F5 & 0.20 & 0.44 & 0.19 & 0.25 & 0.91 & 1.10 & 0.33 & 0.69 & 1.67 & 1.73 & 0.44 & 1.18 \\
F8 & 0.34 & 0.55 & 0.23 & 0.30 & 1.08 & 1.20 & 0.37 & 0.73 & 1.85 & 1.83 & 0.47 & 1.22 \\
G0 & 0.41 & 0.59 & 0.24 & 0.30 & 1.15 & 1.24 & 0.38 & 0.74 & 1.93 & 1.86 & 0.48 & 1.23 \\
G5 & 0.65 & 0.70 & 0.27 & 0.36 & 1.40 & 1.34 & 0.41 & 0.79 & 2.18 & 1.96 & 0.51 & 1.28 \\
G8 & 0.78 & 0.77 & 0.29 & 0.40 & 1.54 & 1.40 & 0.42 & 0.82 & 2.31 & 2.02 & 0.52 & 1.32 \\
K0 & 1.00 & 0.86 & 0.31 & 0.43 & 1.76 & 1.48 & 0.44 & 0.85 & 2.54 & 2.10 & 0.54 & 1.35 \\
K4 & 1.64 & 1.19 & 0.39 & 0.61 & 2.40 & 1.81 & 0.52 & 1.03 & 3.15 & 2.43 & 0.60 & 1.54 \\
M0 & 1.86 & 1.46 & 0.54 & 0.88 & 2.62 & 2.06 & 0.66 & 1.30 & 3.32 & 2.68 & 0.74 & 1.82 \\
M3 & 1.96 & 1.48 & 0.80 & 1.43 & 2.72 & 2.08 & 0.90 & 1.86 & 3.37 & 2.68 & 0.98 & 2.43 \\
M5 & 2.05 & 1.61 & 0.97 & 1.96 & 2.79 & 2.21 & 1.06 & 2.38 & 3.36 & 2.82 & 1.12 & 3.00 \\
M8 & 2.26 & 1.99 & 1.24 & 2.80 & 2.91 & 2.60 & 1.30 & 3.19 & 3.18 & 3.22 & 1.32 & 3.86 \\
\hline
\end{tabular}
\end{table}

\begin{table}[ht!]
\caption{Synthetic colour of selected dwarf stars for $R_V=3.1$}
\begin{tabular}{ccccccccccccc}
\hline
$Spectral$ & \multicolumn{4}{c}{$A_V = 6$} & \multicolumn{4}{c}{$A_V = 8$}  & \multicolumn{4}{c}{$A_V = 10$}  \\
$Type$ & $U_{RGO}-g$ & $g-r$ & $r-H\alpha$ & $r-i$ & $U_{RGO}-g$ & $g-r$ & $r-H\alpha$ & $r-i$ & $U_{RGO}-g$ & $g-r$ & $r-H\alpha$ & $r-i$ \\
\hline
\hline
B0 & 1.09 & 1.75 & 0.48 & 1.15 & 1.89 & 2.35 & 0.58 & 1.56 & 2.62 & 2.97 & 0.65 & 2.10 \\
B3 & 1.53 & 1.84 & 0.48 & 1.21 & 2.30 & 2.44 & 0.58 & 1.61 & 2.97 & 3.05 & 0.64 & 2.17 \\
B5 & 1.67 & 1.88 & 0.47 & 1.24 & 2.42 & 2.48 & 0.57 & 1.64 & 3.06 & 3.09 & 0.63 & 2.20 \\
B8 & 1.81 & 1.92 & 0.46 & 1.26 & 2.56 & 2.51 & 0.56 & 1.66 & 3.16 & 3.13 & 0.62 & 2.22 \\
A0 & 2.10 & 1.96 & 0.43 & 1.28 & 2.82 & 2.55 & 0.52 & 1.69 & 3.36 & 3.16 & 0.58 & 2.24 \\
A2 & 2.21 & 2.00 & 0.43 & 1.32 & 2.92 & 2.59 & 0.52 & 1.73 & 3.43 & 3.21 & 0.58 & 2.29 \\
A5 & 2.31 & 2.06 & 0.44 & 1.36 & 3.01 & 2.64 & 0.53 & 1.76 & 3.49 & 3.26 & 0.59 & 2.32 \\
F0 & 2.40 & 2.21 & 0.51 & 1.44 & 3.10 & 2.80 & 0.60 & 1.84 & 3.50 & 3.41 & 0.65 & 2.41 \\
F5 & 2.42 & 2.32 & 0.56 & 1.51 & 3.12 & 2.90 & 0.64 & 1.91 & 3.47 & 3.52 & 0.69 & 2.49 \\
F8 & 2.61 & 2.41 & 0.59 & 1.55 & 3.27 & 2.99 & 0.66 & 1.95 & 3.54 & 3.60 & 0.71 & 2.52 \\
G0 & 2.69 & 2.44 & 0.59 & 1.56 & 3.34 & 3.02 & 0.67 & 1.95 & 3.57 & 3.63 & 0.72 & 2.53 \\
G5 & 2.93 & 2.53 & 0.62 & 1.60 & 3.52 & 3.11 & 0.69 & 2.00 & 3.62 & 3.72 & 0.74 & 2.58 \\
G8 & 3.06 & 2.59 & 0.63 & 1.64 & 3.61 & 3.16 & 0.70 & 2.03 & 3.61 & 3.78 & 0.75 & 2.61 \\
K0 & 3.26 & 2.67 & 0.64 & 1.67 & 3.74 & 3.24 & 0.72 & 2.06 & 3.62 & 3.86 & 0.76 & 2.64 \\
K4 & 3.74 & 2.98 & 0.70 & 1.84 & 3.88 & 3.54 & 0.77 & 2.23 & 3.41 & 4.17 & 0.81 & 2.84 \\
M0 & 3.79 & 3.21 & 0.83 & 2.10 & 3.71 & 3.76 & 0.89 & 2.49 & 3.09 & 4.39 & 0.92 & 3.11 \\
M3 & 3.70 & 3.22 & 1.05 & 2.66 & 3.46 & 3.77 & 1.10 & 3.05 & 2.77 & 4.40 & 1.13 & 3.72 \\
M5 & 3.51 & 3.37 & 1.18 & 3.18 & 3.09 & 3.93 & 1.21 & 3.57 & 2.33 & 4.55 & 1.23 & 4.30 \\
M8 & 2.89 & 3.78 & 1.36 & 3.97 & 2.24 & 4.36 & 1.37 & 4.34 & 1.42 & 4.99 & 1.36 & 5.11 \\
\hline
\end{tabular}
\end{table}

\begin{table}[ht!]
\caption{Synthetic colour of selected giant stars for $R_V=3.1$}
\begin{tabular}{ccccccccccccc}
\hline
$Spectral$ & \multicolumn{4}{c}{$A_V = 0$} & \multicolumn{4}{c}{$A_V = 2$}  & \multicolumn{4}{c}{$A_V = 4$}  \\
$Type$ & $U_{RGO}-g$ & $g-r$ & $r-H\alpha$ & $r-i$ & $U_{RGO}-g$ & $g-r$ & $r-H\alpha$ & $r-i$ & $U_{RGO}-g$ & $g-r$ & $r-H\alpha$ & $r-i$ \\
\hline
\hline
B0 & -0.93 & -0.10 & 0.10 & -0.07 & -0.22 & 0.58 & 0.26 & 0.37 & 0.53 & 1.24 & 0.39 & 0.85 \\
B2 & -0.78 & -0.06 & 0.21 & -0.03 & -0.08 & 0.62 & 0.37 & 0.41 & 0.67 & 1.28 & 0.50 & 0.90 \\
B5 & -0.54 & -0.04 & 0.15 & -0.03 & 0.15 & 0.64 & 0.31 & 0.41 & 0.88 & 1.29 & 0.43 & 0.90 \\
A0 & -0.09 & -0.00 & 0.03 & -0.01 & 0.58 & 0.67 & 0.19 & 0.43 & 1.29 & 1.32 & 0.31 & 0.91 \\
A5 & 0.17 & 0.11 & 0.09 & 0.09 & 0.84 & 0.78 & 0.24 & 0.54 & 1.55 & 1.42 & 0.36 & 1.03 \\
A7 & 0.26 & 0.27 & 0.06 & 0.12 & 0.93 & 0.93 & 0.21 & 0.56 & 1.64 & 1.58 & 0.33 & 1.05 \\
F0 & 0.32 & 0.34 & 0.15 & 0.21 & 1.01 & 1.00 & 0.29 & 0.64 & 1.73 & 1.64 & 0.41 & 1.14 \\
F2 & 0.33 & 0.44 & 0.17 & 0.20 & 1.05 & 1.09 & 0.32 & 0.63 & 1.80 & 1.72 & 0.43 & 1.12 \\
G5 & 1.06 & 0.84 & 0.30 & 0.43 & 1.82 & 1.47 & 0.44 & 0.85 & 2.59 & 2.08 & 0.53 & 1.35 \\
G8 & 1.25 & 0.89 & 0.31 & 0.44 & 2.01 & 1.51 & 0.44 & 0.87 & 2.78 & 2.12 & 0.54 & 1.37 \\
K0 & 1.33 & 0.91 & 0.31 & 0.46 & 2.09 & 1.52 & 0.45 & 0.88 & 2.85 & 2.13 & 0.54 & 1.38 \\
K3 & 1.92 & 1.15 & 0.36 & 0.57 & 2.68 & 1.76 & 0.49 & 0.99 & 3.41 & 2.36 & 0.58 & 1.49 \\
K5 & 2.53 & 1.32 & 0.40 & 0.66 & 3.28 & 1.92 & 0.52 & 1.08 & 3.94 & 2.52 & 0.61 & 1.59 \\
M0 & 2.72 & 1.43 & 0.51 & 0.93 & 3.46 & 2.02 & 0.62 & 1.35 & 4.05 & 2.62 & 0.70 & 1.87 \\
M3 & 2.78 & 1.50 & 0.60 & 1.17 & 3.50 & 2.09 & 0.71 & 1.59 & 4.05 & 2.69 & 0.78 & 2.13 \\
M5 & 2.33 & 1.50 & 0.75 & 1.83 & 3.05 & 2.10 & 0.84 & 2.24 & 3.59 & 2.72 & 0.89 & 2.82 \\
M8 & 1.18 & 2.18 & 1.10 & 2.76 & 1.87 & 2.81 & 1.15 & 3.17 & 2.36 & 3.49 & 1.16 & 3.83 \\
\hline
\end{tabular}
\end{table}

\begin{table}[ht!]
\caption{Synthetic colour of selected giant stars for $R_V=3.1$}
\begin{tabular}{ccccccccccccc}
\hline
$Spectral$ & \multicolumn{4}{c}{$A_V = 6$} & \multicolumn{4}{c}{$A_V = 8$}  & \multicolumn{4}{c}{$A_V = 10$}  \\
$Type$ & $U_{RGO}-g$ & $g-r$ & $r-H\alpha$ & $r-i$ & $U_{RGO}-g$ & $g-r$ & $r-H\alpha$ & $r-i$ & $U_{RGO}-g$ & $g-r$ & $r-H\alpha$ & $r-i$ \\
\hline
\hline
B0 & 1.30 & 1.86 & 0.52 & 1.22 & 2.09 & 2.45 & 0.61 & 1.62 & 2.79 & 3.07 & 0.68 & 2.17 \\
B2 & 1.43 & 1.90 & 0.62 & 1.26 & 2.21 & 2.50 & 0.71 & 1.66 & 2.88 & 3.11 & 0.78 & 2.22 \\
B5 & 1.63 & 1.91 & 0.56 & 1.26 & 2.39 & 2.51 & 0.65 & 1.66 & 3.03 & 3.12 & 0.71 & 2.22 \\
A0 & 2.01 & 1.93 & 0.44 & 1.27 & 2.74 & 2.52 & 0.53 & 1.68 & 3.30 & 3.14 & 0.59 & 2.23 \\
A5 & 2.26 & 2.03 & 0.49 & 1.37 & 2.97 & 2.62 & 0.57 & 1.78 & 3.45 & 3.24 & 0.63 & 2.34 \\
A7 & 2.36 & 2.18 & 0.45 & 1.40 & 3.05 & 2.77 & 0.54 & 1.80 & 3.47 & 3.38 & 0.59 & 2.36 \\
F0 & 2.46 & 2.23 & 0.52 & 1.47 & 3.14 & 2.82 & 0.61 & 1.87 & 3.51 & 3.44 & 0.66 & 2.45 \\
F2 & 2.55 & 2.30 & 0.55 & 1.46 & 3.23 & 2.88 & 0.63 & 1.86 & 3.56 & 3.49 & 0.69 & 2.43 \\
G5 & 3.31 & 2.65 & 0.64 & 1.67 & 3.77 & 3.22 & 0.71 & 2.06 & 3.66 & 3.83 & 0.76 & 2.64 \\
G8 & 3.48 & 2.69 & 0.64 & 1.68 & 3.88 & 3.25 & 0.71 & 2.07 & 3.68 & 3.87 & 0.76 & 2.66 \\
K0 & 3.54 & 2.70 & 0.65 & 1.69 & 3.91 & 3.26 & 0.72 & 2.08 & 3.67 & 3.88 & 0.76 & 2.67 \\
K3 & 3.97 & 2.91 & 0.67 & 1.80 & 4.05 & 3.47 & 0.74 & 2.18 & 3.54 & 4.09 & 0.78 & 2.77 \\
K5 & 4.30 & 3.05 & 0.70 & 1.89 & 4.09 & 3.60 & 0.76 & 2.27 & 3.42 & 4.22 & 0.80 & 2.87 \\
M0 & 4.26 & 3.15 & 0.78 & 2.15 & 3.91 & 3.71 & 0.84 & 2.53 & 3.18 & 4.33 & 0.87 & 3.15 \\
M3 & 4.16 & 3.22 & 0.86 & 2.39 & 3.73 & 3.78 & 0.91 & 2.77 & 2.97 & 4.40 & 0.93 & 3.40 \\
M5 & 3.70 & 3.26 & 0.96 & 3.04 & 3.27 & 3.83 & 1.00 & 3.42 & 2.51 & 4.47 & 1.00 & 4.09 \\
M8 & 2.46 & 4.01 & 1.21 & 3.94 & 2.02 & 4.59 & 1.22 & 4.32 & 1.25 & 5.26 & 1.18 & 5.09 \\
\hline
\end{tabular}
\end{table}

\end{appendix}

\end{document}